\documentclass[a4paper,fleqn,usenatbib]{mnras}
\usepackage{amsmath}
\usepackage{txfonts}
\usepackage{caption}
\usepackage[T1]{fontenc}
\usepackage{ae,aecompl}
\hypersetup{colorlinks, citecolor=green, filecolor=black, linkcolor=blue, urlcolor=blue }

\usepackage{graphicx}
\usepackage{amssymb}
\usepackage{dblfloatfix}
\usepackage{breqn}
\usepackage{hyperref}
\hypersetup{
    colorlinks=true,
    linkcolor=blue,
    filecolor=magenta,      
    urlcolor=cyan,
}

\title[Physical Spectral-Timing Model]{A Physical Model for the Spectral-Timing Properties of Accreting Black Holes}

\author[Mahmoud \& Done]{
Ra'ad D. Mahmoud$^{1}$\thanks{E-mail: ra'ad.d.mahmoud@durham.ac.uk} \&
Chris Done$^{1}$
\\
$^{1}$Department of Physics, University of Durham, South Road, Durham DH1 3LE\\
}

\date{Accepted XXX. Received YYY; in original form ZZZ}

\pubyear{2018}

\hypersetup{draft}
\begin{document}
\label{firstpage}
\pagerange{\pageref{firstpage}--\pageref{lastpage}}
\maketitle

% Abstract of the paper
\begin{abstract}
We develop new techniques to deconvolve the radial structure of the X-ray emission region in the bright low/hard state of the black hole Cygnus X-1 using both spectral and timing data in the 3-35~keV range. The spectrum at these energies is dominated by Comptonisation rather than the disc, but there is a complex pattern in the time lags between different energy bands and differences in the normalisation and shape in the power spectra of these bands, which clearly shows that the Comptonisation is not produced from a single, homogeneous region. We use a physically based model of density fluctuations propagating through a spectrally inhomogeneous flow, setting the spectral components 
by jointly fitting to the time-averaged and Fourier resolved spectra. The predicted variability in any band is 
modelled analytically in Fourier space so it can be fit directly to the observed power spectra and lags. We find that the best fit model picks out three distinct radii in the flow, each with a 
distinct Compton spectrum. The variability and luminosity produced at these radii is enhanced, 
while propagation of fluctuations from larger radii is suppressed. 
We associate these radii with the disc truncation, the inner edge of the flow, and (more speculatively) the jet launch radius. These distinct radii are most evident where the source is close to a transition between the low/hard and high/soft states. We suggest that the smoother power spectra seen at lower luminosities imply that the source structure is simpler away from the transition. 
\end{abstract}


%%%%%%%%%%%%%%%%%%%%%%%%%%%%%%%%%%%%%%%%%%%%%%%%%%

%%%%%%%%%%%%%%%%% BODY OF PAPER %%%%%%%%%%%%%%%%%%

\section{Introduction}
\label{Introduction}

The overall spectral properties of the low/hard state in black hole
binaries (BHBs) are often interpreted in a truncated disc/hot inner
flow geometry (see e.g. the review by \citealt{DGK07};
hereafter DGK07). In this picture, a cool, optically thick, geometrically thin disc (\citealt{SS73}) transitions at some radius to a hot, optically thin, geometrically thick inner flow which behaves similarly to an advection-dominated accretion flow (ADAF; \citealt{NY95}). A decreasing truncation radius between the cool disc and the hot, inner flow as the mass accretion
rate increases (on timescales of days to weeks) leads to stronger disc
emission, and hence stronger Compton cooling of the hot flow, so that
its spectrum steepens. This also provides a mechanism to explain the
transition to the disc-dominated state when the truncated disc finally
reaches the innermost stable circular orbit (ISCO; \citealt{EMN97}; DGK07).

More recent work has shown in detail how this can also explain the
properties of the fast variability (on timescales of $0.01-10$~s),
including the strong low frequency quasi-periodic oscillation
(QPO). The radius of the thin
truncated disc sets the outer radius of the hot inner flow. Any turbulence in the hot flow will
have a timescale which decreases with radius, so the longest timescale
fast variability is set by the outer radius of the hot flow (\citealt{L97}; \citealt{AU06}); this defines the low-frequency break in the power spectrum. The QPO can be
produced in the same geometry by Lense-Thirring (relativisitic
vertical) precession of the entire hot flow (\citealt{F07}; \citealt{IDF09}), predicting that the iron line centroid energy is modulated on the same quasi-period as the QPO. This characteristic signature of prograde precession 
has been seen in the data in H1743-322 ($3.8\sigma$ detection; \citealt{I16}). 
The decrease in truncation radius with increasing mass accretion rate
which produces the spectral softening then also drives the correlated decrease in low frequency 
break and QPO timescale, as observed 
(\citealt{WvdK99}; \citealt{KWvdK08}; \citealt{ID12b}).

This geometry also offers a qualitative explanation of the more
complex, higher order variability properties. It is well established that BHB lightcurves on fast timescales show
variability in higher energy bands lagging behind that from lower
energy bands (\citealt{MK89}; \citealt{N99}; \citealt{G14}). These lags might be expected in a model where the low energy
band is dominated by the disc, and the high energy band is dominated by
Comptonisation from the hot flow. Fluctuations in the disc will propagate inwards
through the hot flow, but with some lag time corresponding to the viscous timescale
for the fluctuation to drift from the disc inner edge to the innermost parts of the hot flow
which dominate the Comptonised emissivity 
(\citealt{U14};\citealt{R16}). However, such models are unable to 
explain why these lags are seen between different energy bands where the disc emission
is negligible, nor why the lags also depend on the frequency of
variability (\citealt{N99}; \citealt{DM15}; \citealt{M17}). Instead these features can be explained if the hot flow itself is stratified not only in variability timescale, but also in spectral shape, such that the outer regions of the flow are associated with softer Compton spectra than the inner regions (\citealt{KCG01}; \citealt{GDP09}). Slow variability is generated at the largest radii, in the region where the Compton spectrum is softer. These fluctuations have the largest distance to travel (i.e. the longest lag time) before they modulate the hardest spectral inner regions. By comparison, faster variability is produced closer in, so has a shorter distance to
propagate - and hence a shorter lag time - before it modulates the spectrally hardest inner regions (\citealt{MD18a}, hereafter MD18a). If instead the hot flow were homogeneous,
spectra from the outer and inner radii would be the same, so while the fluctuation in the inner radii would be lagged behind those produced in the outer radii, two different energy bands would each sample the same fraction of
initial and lagged emission, so their lightcurves would be the same, and no lag would be observed.

Thus the observed lag between different energy bands which are both
dominated by Comptonisation suggests that the spectral and variability
properties of the hot flow are radially stratified. The observed spectral variability can therefore in principle be used to determine the radial stratification of the hot flow. This goal is
complementary to that of the recent work of \cite{MIvdK18}, where they
focus on using the continuum variability to probe the thin disc structure via reverberation. Here we are interested instead in using the continuum variability to probe the continuum source itself. 

We develop techniques to use the observed spectral-timing information to build a 
model of a flow which is radially stratified in variability, emissivity and energy spectra, developing significantly from our previous work in MD18a.
We use this on some of the best available data, that of the Rossi X-ray Timing Explorer (RXTE) observations of Cyg X-1,
taken before the telemetry limitations which accompanied the antenna failure.
Cygnus X-1 is an ideal source for this study, since there is no obvious QPO to complicate the underlying propagation models, as well as being very bright, resulting in a high signal-to-noise.
We show simultaneous fits to the time averaged spectra, the frequency resolved spectra, and the power spectral densities (PSDs) over different energy bands, together with the
lags between those energy bands.
We show that we can build a self-consistent picture of all these data with a flow composed of two spectral regions, 
with the transition from \textit{soft} to \textit{hard} spectra at very small radii (within $2-4$~$R_g$ of the black hole) 
associated with a large peak in both the variability and the emissivity. 
Alternatively, we prefer a three component spectral model which gives larger radii,
with maximal variability and emissivity 
generated in an intermediary region from $5-6$~$R_g$. While we cannot break all the degeneracies, we show that the data fundamentally require spectrally distinct regions within the Comptonisation zone,
and that the variability and emissivity are jointly enhanced at specific radii. In this picture, the flow consists of a number of bright, turbulent rings rather than being a smooth, self-similar structure. This picture is very different to the original assumption that
variability is dominated by the radially smooth turbulence of the
Magneto-Rotational Instability (MRI), which remains the physical mechanism
behind the angular momentum transport in the flow (e.g. \citealt{BH98}, \citealt{NK09}, as used by \citealt{ID11}, hereafter ID11; \citealt{ID12a}). The bright, turbulent radii we identify must be important in understanding the physics of the flow, and we suggest identifying these with the disc truncation radius, the `bending wave' or nonaxisymmetric tilt-shock radii seen in Magnetohydrodynamic simulations (MHD; \citealt{LOP02}; \citealt{F07}; \citealt{GBFH14}), and/or the jet launch radius.

We note that this complex source structure is required by the distinct timescales picked out 
by the multiple separate maxima (`humps') in the power spectra. A systematic analysis of all the RXTE PCA
data shows that these separate components in the power spectra are especially evident in the
intermediate states, where the source is making a transition between the low/hard and high/soft states 
(\citealt{B02}; \citealt{A05}, \citealt{G14}). Instead, the power spectra are much smoother in the 
dimmer low/hard states, indicating that the X-ray emitting Comptonisation source structure may be 
described by  simpler models where there is a smooth radial gradient in variability timescale, amplitude,
emission spectrum and luminosity in these states (\citealt{R17a}, hereafter R17a). We will apply these spectral-timing 
techniques to data from a dim low/hard state data in a 
subsequent paper (Mahmoud, Done \& De Marco, in preparation). 

\section{Analytic Modelling}
\label{sec:Modelling}
In the following we denote radii as $r=R/R_g$, with gravitational radius $R_g=GM_{BH}/c^2$ and black hole mass $M_{BH}$.

In MD18a we carried out all 
simulations of the mass accretion variability in the hot flow using a
numerical procedure adapted from the method of AU06 and ID11. In that method, the mass accretion rate curve in each annulus, $r_n$, was generated by randomly sampling from a zero-centered Lorentzian with a cut off at the local viscous frequency, $f_{visc}(r)$, 
\vspace*{-3 pt}
\begin{equation}
\label{eq:Lorentzian}
|\tilde{\dot{m}}(r_n, f)|^2 \propto \frac{F_{var}(r_n)^2}{1+[f/f_{visc}(r_n)]^2}.
\end{equation} 
$F_{var}(r_n)$ here is the fractional variability per radial decade at $r_n$, related to the rms variability produced at $r_n$ via $F_{var}(r_n) = \sigma_{rms}(r_n) \sqrt{N_{dec}}$, where $N_{dec}$ is the number of annuli per radial decade. 
The samples were then transformed to the time domain via the Timmer and K\"{o}nig (1995; hereafter TK95) algorithm. These mass accretion rate curves were then lagged and multipled together to simulate propagation through the flow, and the final time series were weight-summed to produce the light curves in each band. 

However these simulations are slow. The basis for a fast procedure to produce the energy-dependent PSDs was also adapted in MD18a from the groundbreaking
analytic formalism of \citealt{IvdK13}, hereafter IvdK13. This analytical method can produce results consistent with those of the TK95 algorithm when spectral constraints are also included. In this work we will use this faster technique to enable us to search through the wide parameter space generated by the multiple components of the model. The updated formalism is described in rigorous detail in Appendix~\ref{Timing Formalism}, but we give an overview of the physical parameters below.

\subsection{Spectral Stratification}
\label{Spectral Stratification}
Fundamentally, our model assumes a changing spectral shape as a function of radius, $F(E,r)$. In the simplest case, it describes a hot flow stratified into two regions, each of which produces a different spectral shape. A \textit{soft} spectral component, $S(E)$, is produced in the outer region, while the inner region produces a \textit{hard} spectral component $H(E)$. In this work, both of these components are formed from Comptonisation
as the RXTE PCA is sensitive only above 3~keV, where the disc emission is neglible. This is different to the 
two component models used by \cite{R17a}, where the lower bandpass of XMM-Newton means that they are sensitive to the direct emission from the disc. Their two component model describes a disc and
homogeneous hot flow, while ours describes an inhomogeneous hot flow. 

$S(E)$, $H(E)$ and their reflected components are determined from direct fitting of a two Compton component model to the broadband spectrum, further details of which are described in Section~\ref{sec:2Comp}. For \textit{soft} and \textit{hard} Comptonisation components (and their reflections) $S(E)$, $H(E)$, $R_S(E)$ and $R_H(E)$, the time-averaged spectrum associated with each annulus is given by
\vspace*{-3 pt}
\begin{equation}
\label{F_r}
\bar{F}(E, r_n)=
\begin{cases}
S(E) + R_S(E) & \text{if}\ r_n > r_{SH}, \\
H(E) + R_H(E) & \text{if}\ r_n < r_{SH}.
\end{cases}
\end{equation}
$r_{SH}$ here is the transition radius between the \textit{soft} and \textit{hard} Comptonisation regions. This is analytically derived from the radial scale, the spectral components $S(E)$ and $H(E)$, and the prescribed emissivity (parameterised in Section~\ref{Correlated Turbulence and Emissivity in a 2-Component Flow}) such that the luminosity ratio between the two direct components, $f_H^S$, matches that of the emissivity, via
\vspace*{-3 pt}
\begin{equation}
\label{r_SH}
f_H^S = \frac {\int_E S(E) dE}{\int_E H(E) dE} = \frac {\int_{r_o}^{r_{SH}} \epsilon(r)
	2\pi r dr }{\int_{r_{SH}}^{r_i} \epsilon(r)  2\pi r dr }.
\end{equation}

\begin{figure}
	\includegraphics[width=\columnwidth]{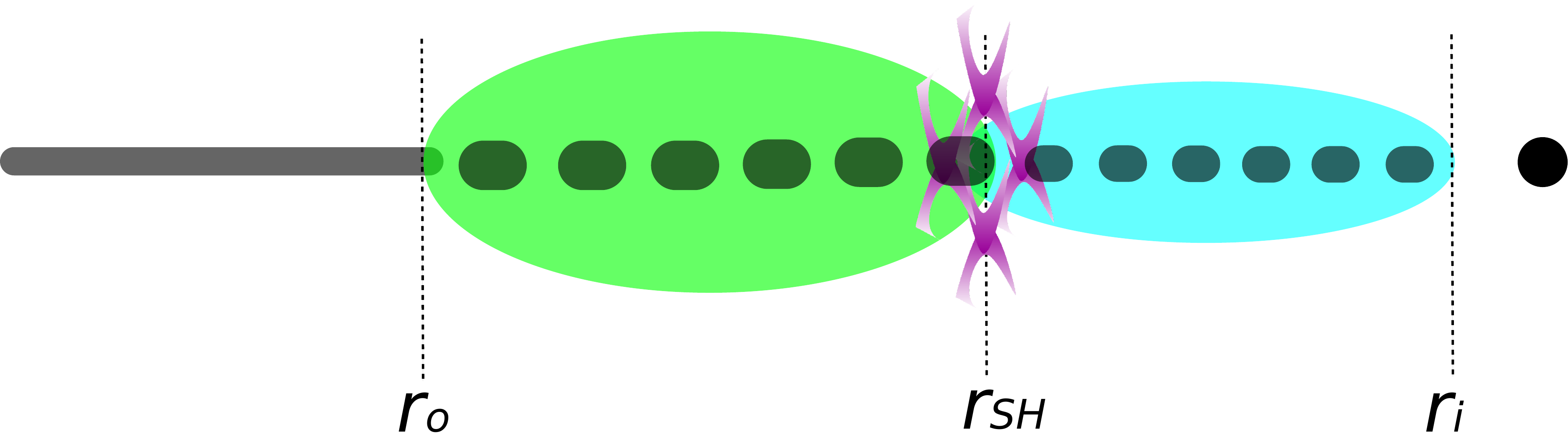}
	\caption{The physical geometry of the flow assumed in the two Compton component model. The dark grey region denotes the thermal, thin disc which does not vary on fast timescales. The green region denotes the fast-varying, spectrally \textit{soft} zone, the cyan region denotes the fast-varying, spectrally \textit{hard} zone. Mass accretes down the flow from the disk truncation radius at $r_o$, through the \textit{soft}-\textit{hard} transition radius at $r_{SH}$, toward the inner flow radius at $r_i$. The purple knots in the flow denote a higher density of magnetic field lines at the spectral transition radius. The grey `blobs' in the flow represent thermal clumps, torn from the disc at the truncation radius, dissipating as they accrete. Clump dissipation, enhancement of the magnetic flux density near the transition radius, or a combination of the two phenomena would result in damping of fluctuations as they propagate from the \textit{soft} to the \textit{hard} region as required by the data, although our model does not distinguish between these mechanisms.}
	\label{fig:TwoGeometry}
\end{figure}

Regarding equation~(\ref{F_r}), we note that our model assumes 
that the observed variability relates only to variation in the normalisation
of each spectral component, not from a changing spectral 
shape. This means we assume that the 
seed photon rate for Comptonisation changes with $\dot{m}$ in the flow,
because the spectral index is set by the 
balance between electron heating from $\dot{m}$ and 
cooling from the seed photons (e.g. \citealt{B99}). 
Any change in the heating and cooling can lead to 
changes in spectral index of the Comptonisation spectrum.
The heating and cooling are expected to vary together if 
the seed photons are predominantly from disc-reprocessed X-ray emission 
(\citealt{HM93}) 
and the light travel time for this reprocessing is fast compared to the propagation time. The latter is fairly well justified for our parameters since the disc truncation radii we derive here from the QPO-low frequency break relation (ID11; Section~\ref{Propagation Speed}) are all <$15~R_g$, equating to a light travel time of order $1$~ms. Of course this is not always the case for AGN (\citealt{GD14}).

However, if the seed photons for the Hard component instead originate from the Soft Compton component, then there will be correlated but lagged variability between the dissipation in the inner regions and its
seed photons caused by the same fluctuation which appeared earlier
in the outer regions. Alternatively (or additionally), changes in spectral shape can be produced even if the seed photons are locally generated, if they arise via cyclo-synchrotron emission. An increase in density gives rise to more local dissipation but fewer seed photons as it increases 
the synchrotron self absorption. This would result in a simultaneous (no lag) anticorrelation of the seed photons with $\dot{m}$. Both cases would lead to strong spectral pivoting (\citealt{V16}).

Nonetheless, spectral pivoting alone (with or without lags) is not likely to explain all of the lags seen in the Cyg X-1 dataset. \cite{MIvdK17} used spectral pivoting
to fit to the complex covariance of this Cyg X-1 data. Here their
focus was to phenomenologically model the continuum lags
in order to extract the reverberation lags from the disc. 
They derived an extremely small inferred inner radius for the disc of $1.5$~$R_g$, but with significant residuals around the iron line. This suggests that their model is using the reverberation lags to compensate for continuum lags which are not modelled with their purely pivoting continuum approach.
We caution that a full picture of the spectral-timing properties of the
flow may require a combination of all these factors, and we will explore this in future work. 

Once $S(E)$, $H(E)$, $R_S(E)$ and $R_H(E)$ are established from the spectrum, the model then solves for parameters which set the radial dependencies of the flow, including
\begin{enumerate}
\item The radial propagation speed $v_r(r)=rf_{visc}(r)$

\item The fractional variability per radial decade, $F_{var}(r)$

\item The total emissivity, $\epsilon(r)$

\item Propagation losses of variability amplitude. This occurs due to either smoothing by the Green's function response of the flow, and/or damping due to macroscopic turbulent processes.
\end{enumerate}
We specify these parameters in detail below, using previous results as the basis to set the minimum number of free parameters required to reproduce the features seen in these data.

\subsection{Propagation Speed}
\label{Propagation Speed}
In simple propagating fluctuations models, the viscous frequency often follows $f_{visc}(r)= B r^{-m} f_{kep}(r)$ where $f_{kep}(r)$ is the Keplerian frequency. This sets the frequency at which fluctuations are generated at $r$ through equation~(\ref{eq:Lorentzian}). Here we assume that this also sets the propagation speed, $v_r(r)$, through $v_r(r) = r f_{visc}(r)$. As we show in Appendix~\ref{PhysicalLags}, the viscous timescale therefore determines not only the peak positions in the PSD, but also the time lag between energy bands.

In MD18a, we used the association of the low frequency QPO with Lense-Thirring precession (see \citealt{I16}) to set $B=0.03$ and $m=0.5$. This was required to reproduce the relation between the low frequency power spectral break and the QPO frequency ($f_{lb}$-$f_{QPO}$; ID11). This relation correlates the low-frequency noise near $f_{lb}$ - that generated in the outermost regions of the flow - with the QPO frequency assuming solid-body precession of the entire hot flow. 

However the low frequency break could instead be set by fluctuations from the inner edge of the disc rather than fluctuations in the hot flow itself. In this work, we therefore assume that $B=0.03$ and $m=0.5$ in the outer part of the flow, but we also explore whether the inner flow has a distinct propagation speed, as might be expected from the transition between the thin disc and the hot flow (\citealt{HR17}). Where the spectral shape switches from \textit{soft} to \textit{hard}, the physical process resulting in the different spectrum may also be associated with a different viscosity form. We therefore parameterise the viscous timescale as
\begin{equation}
\label{eq:2visc_2comp}
f_{visc} = 
\begin{cases}
B_{S} r^{-m_S} f_{kep}(r) & \text{if } r\geq r_{SH} \\
B_{H} r^{-m_H} f_{kep}(r) & \text{if } r<r_{SH},
\end{cases}
\end{equation}
where $B_S$, $m_S$, $B_H$ and $m_H$ are model parameters for the viscosity in the \textit{soft} and \textit{hard} regions. Regardless of the internal variability processes of the flow at $r << r_o$, the $f_{lb}$-$f_{QPO}$ relation should hold, so we fix $B_S=0.03$ and $m_S=0.5$. However we do allow $B_H$ and $m_H$ to vary, as the inner-region viscosity is not constrained by the QPO.

We note that some previous works (e.g. \citealt{ID12a}, \citealt{R16}, R17a) have instead used a viscous frequency profile for the flow which smoothly varies with surface density, this form being inferred from MHD simulations. However those works also assume smooth, MRI-driven turbulence and emission profiles, as these were satisfactory to explain their data given the lack of spectral constraints. In the new picture we are exploring - of quasi-limited regions enhancing the variability, and in particular of the thin disc being shredded in the \textit{soft} region - it becomes inconsistent to think of the viscous timescale as being a smooth function of radius. This motivates the broken profile of equation~(\ref{eq:2visc_2comp}).

\subsection{Correlated Turbulence and Emissivity in a 2-Component Flow}
\label{Correlated Turbulence and Emissivity in a 2-Component Flow}

MD18a principally showed that the `bumpy' power spectra seen in the brighter
low/hard states of Cyg X-1 and other sources (\citealt{CGR01}; \citealt{P03}; \citealt{A08}; \citealt{T11}; \citealt{M17}; GX 339-4: \citealt{N00}) cannot be matched by self-similar
turbulence where $F_{var}(r)$ is constant. Regions of enhanced variability over a very small range in radii are required in order to produce an excess of power over a limited frequency range. These may be
physically associated with the transition from the thin disc to the hot
flow, nonaxisymmetric shocks
at the inner radius of the flow if this is tilted with respect to the
black hole spin (\citealt{HBF12}; \citealt{GBFH14}) or the jet/flow interaction.

The PSDs in our data show three distinct bumps (the shaded regions in e.g. Fig.~\ref{fig:MD17003}a). In an effort to reproduce the clearly complex structure seen in the variability, we would like a phenomenological model for the turbulence as a function of radius which does not require an excess of free parameters. We therefore parameterise
$F_{var}(r)$ as a sum of three gaussians, with width, $\sigma^{en}_{j}$,
radial position, $r^{en}_{j}$ and amplitude, 
$A_{j}$, giving
\begin{equation}
\label{eq:newFvar}
F_{var}(r) = \sum_{j=1}^{3} A_{j} e^{-\frac{r-r^{en}_{j}}{2\sigma^{en}_{j}}}.
\end{equation}
One very natural source of this variability is at the
truncation radius, where the interaction between the between the Keplerian disc and sub-Keplerian flow is likely to be highly unstable. We therefore associate the outermost gaussian in $F_{var}(r)$ with the truncation radius so that $r^{en}_1 = r_o$. We also impose joint constraints on the other parameters in equation~(\ref{eq:newFvar}). The characteristic radii of the two inner gaussians are always limited to be less than that of the adjacent outer one, i.e. $r^{en}_2 < r^{en}_1 (= r_o)$ and $r^{en}_3$ < $r^{en}_2$. $r^{en}_3$ also cannot fall below the inner radius of the flow, so that $r^{en}_3 > r_i$. We force positive-definite resulting lightcurves by ensuring that the
rms variability generated in any region has $0 < \sigma_{rms}(r) < 0.3$. Models which go above this 
limit are flagged as infinitely bad in terms of goodness of fit (see Section~\ref{fitting}). Since $\sigma_{rms}(r) \propto F_{var}(r)$, this implies joint upper boundaries on the gaussian widths and amplitudes, $\sigma^{en}_{j}$ and $A_{j}$.

Incorporating equation~(\ref{eq:newFvar}) into the model allows it the
freedom to significantly vary the amount of turbulence within the flow as a function of position. The physical pictures which can be inferred from the fits can therefore range from the traditional propagating fluctuations model, with broad, low amplitude Lorentzians producing constant variability at each radial decade, to the case of certain processes dominating the variability at distinct radii, potentially independent of any smooth MRI-driven turbulence.

However, to make an impact on the power spectra, regions of 
enhanced variability should form a 
significant contribution to the 
resulting lightcurves. This requires that they modulate a large fraction of the total luminosity. 
It is difficult to do this with a smooth emissivity (as assumed in MD18a) as this limits the 
contribution to the lightcurve from any small range of radii. The impact of a turbulent region in the
lightcurve is a product of the variabilty power and the integrated emissivity over the radial range where the enhanced variability appears. With the smooth emissivity assumption, the additional variability required to fit the data is so
large that the assumption of linearity in the generated
lightcurves is broken (MD18a). This results in the lightcurves becoming negative, which is unphysical!

We now link the turbulence to the emissivity of the flow, such that the more variable regions produce a larger proportion of the 
emission. This is motivated not only by the demands of the data, but also by the findings of MHD simulations which show that turbulence is inherently dissipative (see e.g. \citealt{B13}). We parameterise the radial dependence of the emissivity with a sum of three Gaussians, tied to the same radii and width as for the turbulence above, but with free normalisations. We also assume that this occurs on thebackground of a smooth $\epsilon(r)$ with index $\gamma$ in order to encode the joint contributions from the turbulence and gravitational energy release to the energetic dissipation. We then have
\begin{equation}
\label{eq:newemiss}
\epsilon(r) \propto r^{-\gamma} \left( e^{-\frac{r-r^{en}_{  1}}{2\sigma^{en}_{ 1}}} + Z_{ 2} e^{-\frac{r-r^{en}_{  2}}{2\sigma^{en}_{ 2}}} + Z_{ 3} e^{-\frac{r-r^{en}_{  3}}{2\sigma^{en}_{ 3}}}\right),
\end{equation}
where $Z_{ 2}$ and $Z_{ 3}$ are the relative amplitudes of the second and third Gaussians to that of of the outermost at $r_o$. Through equation~\ref{eq:newemiss}, the model can also encompass interference-based interpretations of the double-humped power spectra, where the correlated, lagged variability between Comptonised photons from the outer and inner regions results in suppressed power in the range of frequencies for which the lag is half of the fluctuation period (\citealt{V16}). On the other hand if interference is not significant, i.e. if the propagation delay is too short, damping will be required to suppress the propagated power and produce dips in the power spectra (see Section~\ref{SmoothingDamping}).

To calculate light curves in different energy bands from our simulation, we weight the mass accretion rate at each annulus $r_n$ by the product of the emissivity at $r_n$, and the integrated spectral components of equation~(\ref{F_r}), folded with the detector response and interstellar absorption (see equation~\ref{eq:weights}). Including the detector response and interstellar absorption ensures that the simulated data is weighted in the same way as the observations.

\begin{figure}
	\includegraphics[width=\columnwidth]{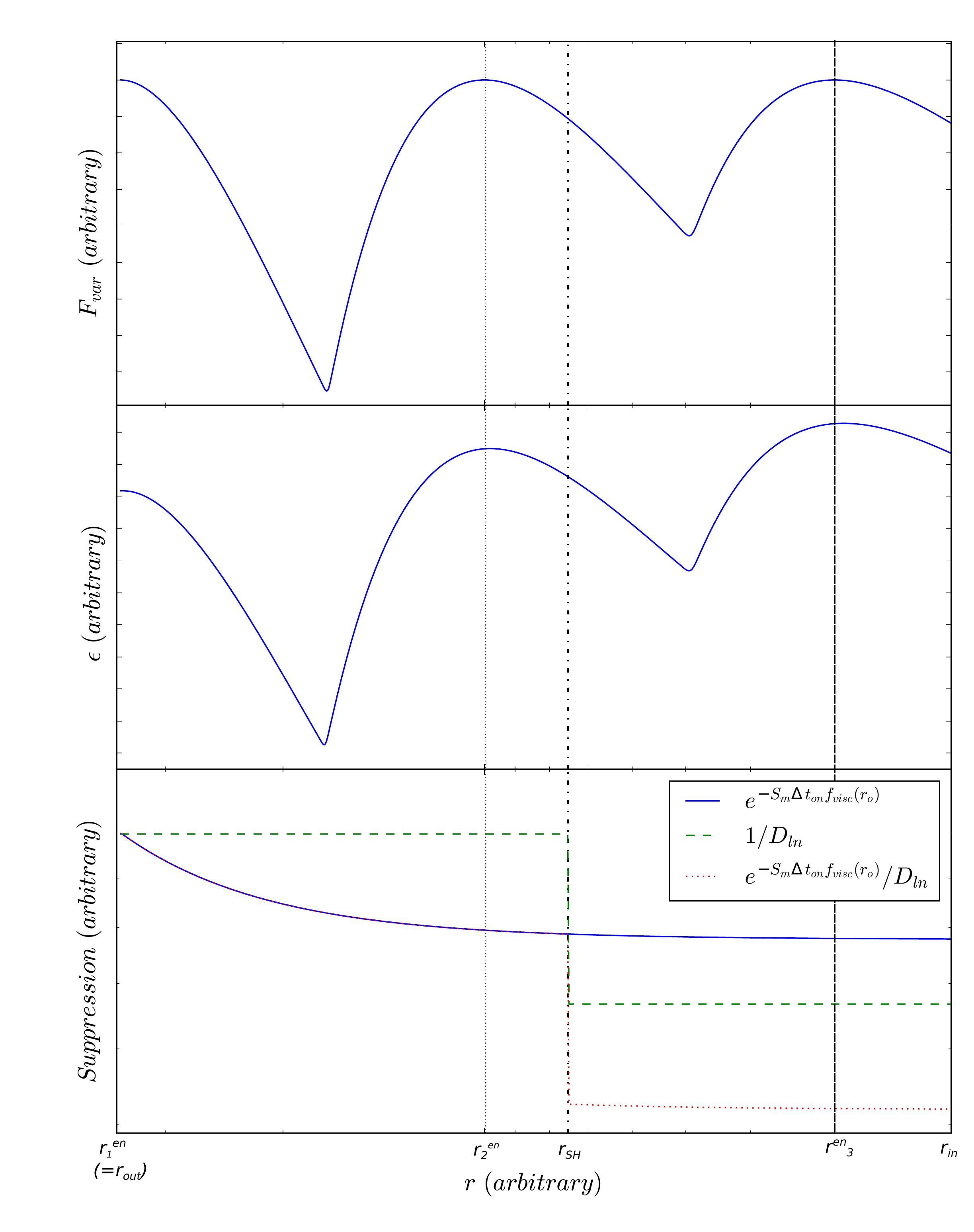}
	\caption{Top panel (a): generic schematic of the fractional variability profile used in the model (blue solid line). Middle panel (b): generic schematic of the emissivity profile used in the model (blue solid line). Comparing panels (a) and (b), we illustrate that the location of the humps in variability correspond to those in the emissivity, from the assumption that turbulence ties to energetic dissipation. Bottom panel (c): generic schematic of the smoothing/damping profiles, denoting the effect of damping (green, dashed line), an example of smoothing (blue, solid line), and the product of these as used in the model (orange, dotted line). Note that we only show an \textit{example} of smoothing at a single frequency propagating from the truncation radius (that of $f_{visc}(r_o)$), while smoothing is actually a two-dimensional function of frequency and propagation time. The black dot-dashed line denotes the spectral transition radius, $r_{SH}$.}
	\label{fig:Generic_Fvar_em_suppression}
\end{figure}

In Fig.~\ref{fig:Generic_Fvar_em_suppression}(a) \& (b) we show the schematic for the revised
$F_{var}(r)$ and $\epsilon(r)$ profiles with arbitrary normalisation
on both, illustrating how these profiles are now associated
in terms of the positions and radial scales of the enhanced regions.

\subsection{Smoothing and Damping}
\label{SmoothingDamping}

Another key feature often seen in the Cyg X-1 low/hard states which previous models have been unable to replicate is the suppression of low-frequency correlated variability associated with lower energy bands compared to higher energy bands (e.g. \citealt{G14}, \citealt{R17b}; MD18a). This feature is also seen in other BHBs including SWIFT J1753.5-0127 and GX 339-4 (\citealt{WU09}), and so is likely generic to the low/hard state paradigm. For similar Low and High energy power spectral shapes (i.e. no preferential suppression of certain frequencies between bands), the timing properties can be jointly fit using simple propagating fluctuations, but when the power spectral shapes become distinct at low frequencies, the lags in each band become dominated by different spectral components and joint fits fail (\citealt{R17a}). This implies that the variability present in the spectrally softer regions is not completely propagated down to modulate the harder regions. Yet the fact that there are lags between Low and High bands on the timescale of the outer parts of the flow means that at
least \textit{some} of the variability does propagate. The variations in the \textit{soft} region must therefore map onto the inner-region variability after propagation, although with smaller amplitude and a time delay. 

Our model includes two distinct methods of variability suppression. First, there can be 
diffusion of fluctuations as they propagate through the flow due to the Green's function response (e.g. R17a). This acts to preferentially smooth out the fluctuations which have propagated over a 
larger distance compared to their wavelength. 
We follow R17a and model this as an exponetial decay such that all fluctuations of frequency $f$ 
which propagate from $r_l$ to $r_n$ (which takes timescale $\Delta t_{ln}$) are suppressed
by a continuous factor $exp(-S_m  \Delta t_{ln} f)$. Here
\begin{equation}
\Delta t_{ln} = \sum_{k = l}^{n-1} d t_k = \sum_{k = l}^{n-1}\frac{dr_k}{r_k} t_{visc}(r_k)  = dlog(r_k) \sum_{k = l}^{n-1} t_{visc}(r_k),
\end{equation}

However, our model fundamentally includes regions of enhanced variability/emissivity. The specific processes which generate this additional power could also suppress the propagating fluctuations at the associated radii. In our nomenclature, we will refer to the process of suppressing power at all frequencies uniformly at specific radii as `damping', distinct from the diffusive `smoothing' process discussed above, which instead has a frequency/travel time dependence. We tie our damping effect to the radius at which the spectral shape changes from \textit{soft} to \textit{hard}. There are two proposed mechanisms by which such damping could occur, the first being an increased density of magnetic field lines at the transition radius. This buildup of flux density could be associated with the jet collimation, or warps induced by the flow tilt with respect to the disc plane. Such a process would be highly disruptive to incoming mass accretion rate fluctuations from $r> r_{SH}$. The purple knots in the flow in Fig.~\ref{fig:TwoGeometry} represent this concentration of flux density at the spectral transition radius. An alternative, possibly concurrent mechanism which would result in fluctuation damping could be the dissipation of thermal clumps propagating through the flow. Physically, the spectral transition requires a change in seed photon availablity, and these seed photons in the \textit{soft} region could be from clumps torn from the inner edge of the truncated disc by the truncation process. These clumps would produce copious, highly variable, seed
photons for the \textit{soft} region of the flow, making the characteristic soft, highly variable spectrum (see also \citealt{U11}; \citealt{G14}). However as it accretes, a given clump would likely shrink due to its outer layers thermalising with the surrounding medium, and being stripped away by the MRI; this would also result in seed photon starvation of the inner region, explaining the observed hardening of the spectral shape. The grey `blobs' in the flow in Fig.~\ref{fig:TwoGeometry} represent these clumps dissipating as they accrete. When considering this possibility however, we bear in mind the work of \cite{PVZ18} which showed that synchrotron emission may be the preferred method source of seed photons in the inner flow, although incorporating this into the model would invoke the aforementioned complexity of spectral pivoting.

Applying a generic mechanism which can encompass both magnetically-driven and clump evaporation-driven damping, we suppress all the propagated variabilty by a single factor $D_{SH}$ at the spectral transition radius, $r_{SH}$. In the context of fluctuations propagating from $r_l$ to $r_n$, these fluctuations are therefore damped by 
\begin{equation}
\label{2damp}
D_{ln} =
\begin{cases}
D_{SH} & \text{if } r_l > r_{SH} > r_n, \\
1 & \text{otherwise}.
\end{cases}
\end{equation}

In Fig.~\ref{fig:Generic_Fvar_em_suppression}(c) we show an example of these smoothing and damping effects separately, and in combination. Note that the smoothing in this figure is an example for a fixed frequency, $f=f_{visc}(r_o)$, and in truth smoothing acts as a function of frequency.

\section{Timing Fit Procedure}
\label{fitting}
Producing a fully generalised model which fits the spectra and timing
properties simultaneously would be both technically complex and
computationally prohibitive. Furthermore, in a parameter space with
dimensions corresponding to the free parameters for both the spectral
and timing fits, degeneracy and correlation between parameters would be a significant concern. With this in mind, we fit first to the energy spectra, and then to the
energy-dependent timing properties, to demonstrate the possible achievements and limitations of the model. Unlike the slower TK95 method used in MD18a, the fast analytic prescription we use allows us to obtain parameter space minima using the Markov Chain Monte Carlo (MCMC) method via the {\sc{python}} package, {\sc{emcee}} (\citealt{DFM13}). This allows a much more thorough exploration of the parameter space than in MD18a, systematically testing the limits of each model. In all analytic model fits, we assume that Cyg X-1 has a black hole of mass $M_{BH} = 15 M_{\odot}$, and a dimensionless spin parameter of $a^*\sim$~0.85 when calculating $f_{kep}(r)$ (\citealt{T14}; \citealt{K17}). All spectral-timing fits use $N_r=70$ logarithmically spaced radial bins, which is a compromise between computational cost and spatial resolution. We simulate on a time binning of $dt = 2^{-6}$~s to match the Nyquist frequency of our data. 

To find the timing properties of the data, we extract lightcurves using {\sc{saextrct}} in three energy bands: Low (3.13-4.98~keV), Intermediate (9.94-20.09~keV) and High (20.09-34.61~keV). Selecting the boundaries of the Low and Intermediate bands ensures that contamination of the timing signal from the iron line at $6.2$~keV is avoided. Since the PCA response declines rapidly below $3$~keV and HEXTE becomes unreliable above $35$~keV, these bands span as large an energy range as possible for these data without, significantly sacrificing signal-to-noise. These lightcurves are found by co-addition of three consecutive observations of Cyg X-1 (ObsIDs: 10238-01-08-00, 10238-01-07-000, 10238-01-07-00, hereafter Obs. 1-3), where the similar spectra and hardness ratios of these observations justify their co-addition (see MD18a). Obs. 1-3 all consist of 64 energy bins across the entire RXTE PCA energy bandpass (standard channels 0-249; {\tt{B_16ms_64M_0_249}} configuration) giving moderate spectral resolution in the 3-10 keV band near the iron K$\alpha$ line. We calculate the noise-subtracted power spectra and time lags of the data by ensemble averaging over 174 intervals, each containing $2^{13}$ time bins of $2^{-6}$~s length. The power spectra and time lags are then rebinned geometrically such that the number of points in bin $x$ adheres to $N_p(x) \leq 1.11^x$ (\citealt{vdK89}), with at least a single point in each frequency bin. This produces fully binned power spectra in each band, $P_{i}(f)$, and lags between bands, $\tau_{ij}(f)$, where $i, j = 1\ldots N$ and $N$ is the number of distinct energy bands. These statistics have corresponding errors $\Delta P_i(f)$, $\Delta \tau_{ij}(f)$. Since each measured frequency is first ensemble averaged over 174 separate intervals, this far exceeds the prerequisite number of points required for the errors to converge to Gaussian at all frequencies (\citealt{PL93}), and so chi-squared fitting is appropriate for this data.

When fitting, we wish to maximise our signal-to-noise ratio in the time lag. Due to the closer proximity of the High and Intermediate (and Intermediate and Low) bands, the propagation lag between these bands is inherently smaller, while their errors are the same as that of the High-Low. Particularly in the case of the Intermediate-High band cross-spectrum, the average lag reduces to $\sim 10^{-3}$~s, of the order which would be expected for reverberation lags given the geometry of the system. Since the model does not include the reverberation lag here, fitting to these lag spectra would therefore skew the fit statistic towards an underpredicted lag, breaking the fit.

To simultaneously fit to the power spectra and time lags, we therefore reduce the sum of the chi-squared values for the power spectra in the Low, Intermediate and High bands, and a weighted form of the High-Low band time lags, since the High-Low lag gives us the best signal to noise ratio, thus being the most reliable diagnostic of the propagation lag. In particular we weight the lag by a factor of three against the power spectra. Selection of a factor three ensures an equal balance in fit preference between the overall power spectral and lag statistics.
The statistic we reduce is therefore 
\begin{align*}
\chi^2 = \sum^J_{j=1} \left\{ \frac{[P^{\,mod}_{Lo}(f_j)-P^{\,obs}_{Lo}(f_j)]^2}{\Delta P^{\,obs}_{Lo}(f_j)^2} + \frac{[P^{\,mod}_{Int}(f_j)-P^{\,obs}_{Int}(f_j)]^2}{\Delta P^{\,obs}_{Int}(f_j)^2} + \right. \\ \left. \frac{[P^{\,mod}_{Hi}(f_j)-P^{\,obs}_{Hi}(f_j)]^2}{\Delta P^{\,obs}_{Hi}(f_j)^2} + 3\frac{ [\tau^{\,mod}_{LH}(f_j)-\tau^{\,obs}_{LH}(f_j)]^2}{\Delta \tau^{\,obs}_{LH}(f_j)^2}\right\},
\end{align*}
where the superscripts $mod$ and $obs$ refer to the model and observational statistics respectively, and subscripts $Lo$, $Int$ and $Hi$ refer to the Low, Intermediate and High energy bands respectively.

\section{Spectral Fits: Two Compton Components}
\label{sec:2Comp}

\begin{figure}
	\includegraphics[width=\columnwidth]{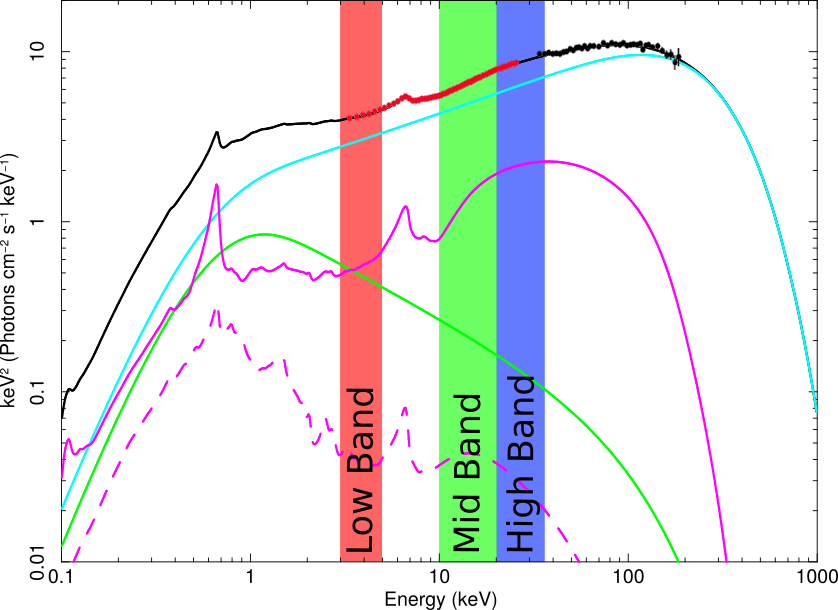}
	\caption{Two component spectral decomposition of Obs. 1, denoted model spectral model \textbf{\textit{2C}}. Lines show the total energy spectrum (black solid), the \textit{hard} Compton component ($H(E)$, cyan solid), the \textit{soft} Compton component ($S(E)$, green solid), the truncated disc reflection from the \textit{hard} component ($R_H(E)$, magenta dashed), and the reflection from the \textit{soft} component ($R_S(E)$, magenta solid). Filled circles show the PCA (red) and HEXTE (black) data. The red, green and blue bands denote the Low (3.13-4.98~keV), Intermediate (9.94-20.09~keV) and High (20.09-34.61~keV) energy ranges respectively. Systematic errors on model and data have been updated leading to very different spectral shape from that of MD18a which fit the same data with the same model.}
	\label{fig:MD17}
\end{figure}

\begin{table*}
	\centering
	\begin{tabular}{ccccccccccc}
		\hline
		Spectrum & $\Gamma_S$ & $kT_{e, S}$$^\dagger$ & $n_S$ & $\Gamma_H$ & $kT_{e, H}$$^\dagger$ & $n_H$ & $\left( \frac{\Omega}{2\pi}\right)_S$ $=\left( \frac{\Omega}{2\pi}\right)_H$ &log($x_i$) & $\chi^2/dof$  \tabularnewline
		& & (keV) & & & (keV) & & & & &  \tabularnewline
		\hline
		\vspace{+3pt}
		Total & $2.6^{+0.4}_{-0.3}$  & $170^{+50}_{-20}$ & $0.82^{+0.05}_{-0.11}$ & $1.65\pm{0.02}$ & $170^{+50}_{-20}$ & $1.7^{+0.1}_{-0.2}$ & $-0.28\pm 0.01$ & $3.00^{+0.01}_{-0.03}$ & $102.1/91$   \\
		\hline
		\multicolumn{11}{l}{\textsuperscript{$\dagger$}\footnotesize{These are tied.}}
	\end{tabular}
	\caption[caption]{Fitting parameters for spectral model \textbf{\textit{2C}}, described by {\tt{tbnew_gas * (nthcomp + nthcomp + kdblur * xilconv * (nthcomp + nthcomp))}}. This shows fit parameters close to that of MD18a, only now with separate reflection from each Compton component. $n_m$ and $\left(\frac{\Omega}{2\pi}\right)_m$ denote the normalisation and reflection fractions on Compton component $m$. Normalisation values are in standard units of $photons$ $s^{-1}$ $cm^{-2}$ at $1$~keV. Uncertainties are quoted at the $1\sigma$ confidence level. The associated spectrum is shown in Fig.~\ref{fig:MD17}.}
	\label{tab:2specparams}
\end{table*}

\begin{figure}
	\includegraphics[width=\columnwidth]{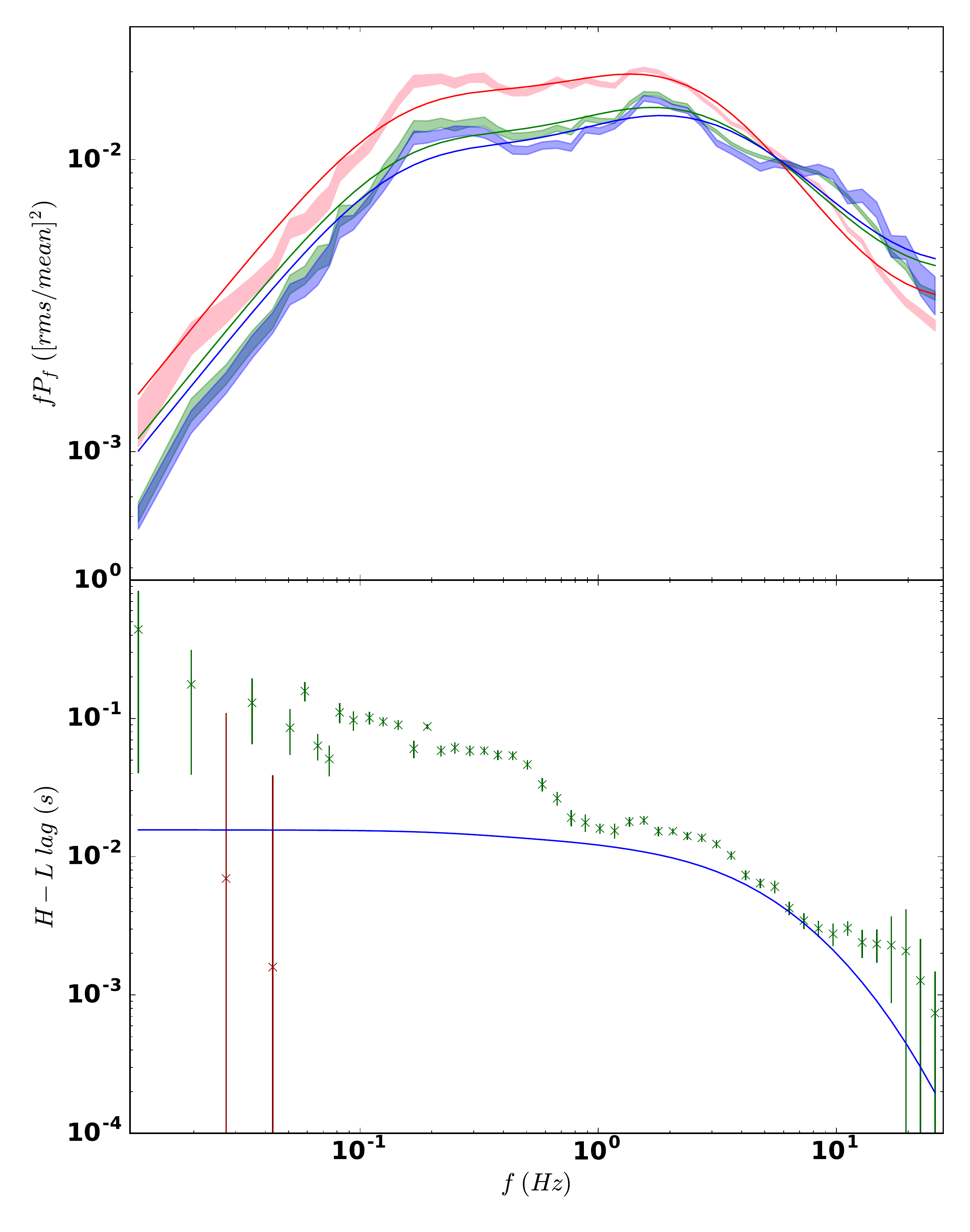}
	\caption{Timing fit using spectral model \textbf{\textit{2C}}. Top panel (a): High, Intermediate \& Low band PSDs. The shaded regions are the 1$\sigma$ error regions of the Low (pink), Intermediate (green) and High (blue) energy bands from the data. The solid lines show the Low (red), Intermediate (green) and High (blue) energy model outputs. Bottom panel (b): Crosses denote the time lags between the High and Low bands for the data. Green crosses indicate the High band lagging the Low band. Red crosses indicate the Low band lagging the High band. The blue solid line denotes the model output.}
	\label{fig:MD17003}
\end{figure}

\begin{figure}
	\includegraphics[width=\columnwidth]{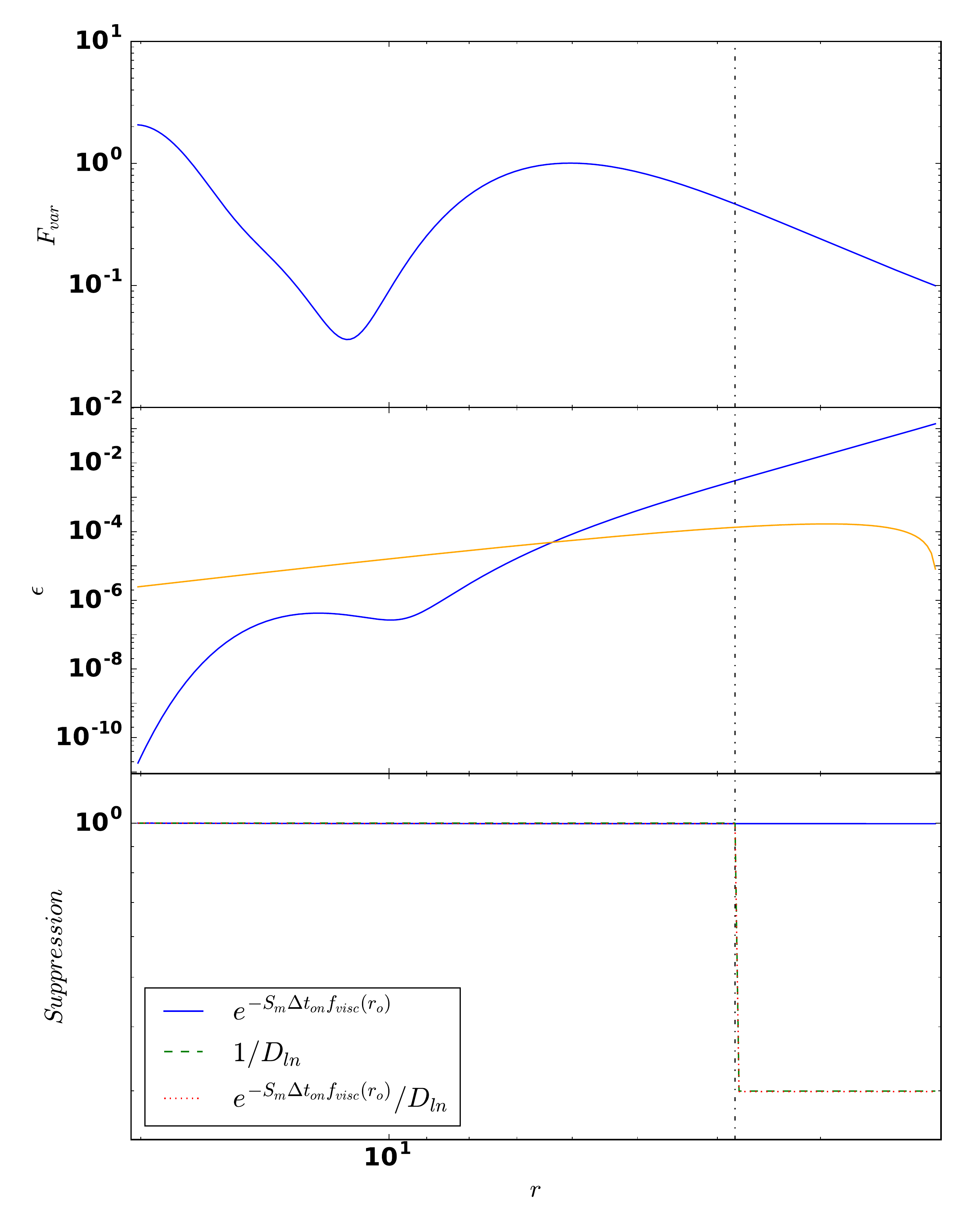}
	\caption{Top panel (a): Fractional variability ($F_{var}$) profile found for fit of Fig.~\ref{fig:MD17003}. Colours and linestyles as in Fig.~\ref{fig:Generic_Fvar_em_suppression}(a). Middle panel (b): Emissivity ($\epsilon$) profile found for fit of Fig.~\ref{fig:MD17003}. Orange solid line denotes Novikov-Thorne-type $\epsilon (r) \propto r^{-3} \left(1-\sqrt{r_i/r}\right)$ profile for comparison. Other colours and linestyles as in Fig.~\ref{fig:Generic_Fvar_em_suppression}(b). Bottom panel (c): Smoothing/damping profile found for fit of Fig.~\ref{fig:MD17003}. Colours and linestyles as in Fig.~\ref{fig:Generic_Fvar_em_suppression}(c). }
	\label{fig:MD17003_fvar_emiss_suppression}
\end{figure}

In MD18a we fitted the time averaged SED from Obs. 1 alone in {\sc{xspec}} (version 12.9.1; \citealt{ABH96}). The model consisted of two Comptonisation components described by {\tt{tbabs * (nthcomp + nthcomp)}} (\citealt{ZJM96}), and their combined reflection, {\tt{tbabs * (kdblur * xilconv * twocomp)}}, where {\tt{twocomp}} is a local model which adds the Comptonisation components together. This fit was motivated by model simplicity, and by similar successful fits to Cyg X-1 spectra (\citealt{G97}; \citealt{DS01}; \citealt{M08}; \citealt{B17}). In MD18a this fit assumed 0.5\% errors on the model alone, and no systematic errors on the data. 

For the first spectral fit of this paper, we make two key revisions to this approach. First we use a slightly updated version of the MD18a model, instead fitting {\tt{tbnew_gas * (nthcomp + nthcomp + kdblur * xilconv * (nthcomp + nthcomp))}}\footnote{{\tt{tnew_gas}} is the new, faster version of {\tt{tbabs}} (\citealt{WAM00}), and can be found here: \\ pulsar.sternwarte.uni-erlangen.de/wilms/research/tbabs.}, where the reflected components is now separated into its \textit{soft} and \textit{hard} components to be consistent with the revised formalism of equation~(\ref{F_r}). Second, we now add 1\% systematic errors to the data only in the PCA range, and assume zero systematic error on the model, as opposed to the zero data-systematic, 0.5\% model systematic used in MD18a. This decision is made for consistency with later fits in this paper, which incorporate Fourier-resolved spectra, using 1\% PCA systematics as in \cite{AD18}. Like MD18a however, we have followed \cite{M08} who fix the seed photon temperature for both Compton components to that of the disc at $0.2$~keV, and tie the electron temperatures of both components. We also fix the galactic column density to $N_h = 6 \times 10^{21}$~cm$^{-2}$ in this and all other fits. Unlike the timing model, we fit the spectrum to Obs. 1 only, as co-addition of spectral data can result in artefacts.

In Fig.~\ref{fig:MD17} we show the broadband spectral fit. The dominant \textit{soft} (green) and \textit{hard} (cyan) Compton components are produced from the outer and inner regions of the flow respectively. Also included is the reflection from the \textit{soft} component (magenta solid), and from the \textit{hard} component (magenta dashed), but we do not include the intrinsic or reprocessed disc emission as the energy of this is too low to make a significant contribution to the RXTE data above 3~keV. We denote this as spectral model \textbf{\textit{2C}} (two component), with full parameters detailed in Table~\ref{tab:2specparams}. On comparison to the spectral fit of MD18a which found \textit{soft} and \textit{hard} spectral indices of 1.8 and 1.25 respectively, it is clear that the spectral picture we have found here is \textit{dramatically} different in shape, despite the data being identical and the model having no additional complexity. The discrepancy between them therefore arises entirely from the systematic errors assumed. This exemplifies how susceptible spectral data alone are to degeneracy, when no external constraints are imposed from e.g. the timing properties (see also \citealt{B17}). Due to the constraints on the seed photon and electron temperatures, the quality of spectral fit \textbf{\textit{2C}} is fairly poor ($\chi^2_{\nu}=102.1/91$) although in Section~\ref{Fourier Resolved - 2 component} we will look at how releasing these constraints can in fact lead to a better fit to the total and Fourier-resolved spectra.

Fig.~\ref{fig:MD17003} shows the optimal joint fit to the PSDs of all three energy bands, and to the High-Low band time lags obtained using spectral model \textbf{\textit{2C}}. The parameters of this fit are presented in Table~\ref{tab:params}, along with parameters for all other spectrally-constrained timing fits shown in the body of this paper. The model fits reasonably well to the power spectral properties, matching the higher normalisation of 
the Low energy band at low frequencies compared to the Intermediate and High, and the switch to lower
normalisation above 8~Hz. However, the model has much less of the 
double hump structure below $5$~Hz than is seen in the data in all energy bands. This is despite the 
freedom to add regions with enhanced variability, emissivity and damping/smoothing, 
with the best fit radial profiles shown in 
Fig.~\ref{fig:MD17003_fvar_emiss_suppression}. This has 
significant variability in the outer parts of the flow with a hump in $F_{var}(r)$ at $r_o$. This variability then propagates down through the \textit{soft} region, with most contribution to the Low-energy light curve coming from the inner parts of the \textit{soft} region due to the centrally peaked emission profile. The significant damping 
(dashed line in the  lower panel)  between the \textit{soft} and \textit{hard} regions (D_{SH} = 3.3) has then resulted in the correct power spectral hierarchy, with the \textit{hard} component modulated by only a small part of the low frequency variability propagated from the \textit{soft} region.There is no significant smoothing associated with the flow (solid blue line in the lower panel). 

However, this best fit solution completely underpredicts the lag. By having a centrally peaked emission profile which does not contribute much beyond $5$~$R_g$, the \textit{soft} variability becomes characterised almost entirely by that near its inner edge close to $r_{SH}$. This results in a small `characteristic' propagation time from the \textit{soft} to \textit{hard} region, far underpredicting the long lags seen in the data (see equation~\ref{eq:raves} and Fig.~\ref{fig:TwoLag}).

This tension between the power spectra and lag results are likely due to fundamental problems with spectral model \textbf{\textit{2C}}, as the weightings from the spectral fit dictate the contributions each energy band will see from each Comptonisation region. Modification of these spectral weights through an alternative spectral fit can therefore drastically alter the shape of the parameter space explored by the timing model. We will therefore now examine what our timing model can achieve when fitting only to the power spectra and time lags, with no prior fit to the energy spectra.

\subsection{No Constraints from the Spectrum}
\label{Weights_Free}

Reworking the model to require no prior spectral fit can be done quite simply, by rewriting equation~(\ref{eq:weights}). This equation describes the weighting of the mass accretion rate curve from annulus $r_n$ when calculating the light curve in energy band, $i$. Recasting the energy-dependent part of this equation, we get
\begin{align}
\label{eq:free_weights1}
w_{n}^{\,i} &= \frac{\epsilon(r_n) r_n dr_n}{\sum\limits_{region} {\epsilon(r_n) r_n dr_n}}\int\displaylimits_{E = E_{i}^{min}}^{E_{i}^{max}} \bar{F}(E, r_n)A_{eff}(E)e^{-N_H(E)\sigma_T} dE \nonumber\\
&=\frac{\epsilon(r_n) r_n dr_n}{\sum\limits_{region} {\epsilon(r_n) r_n dr_n}} L_{k(r_n)}^{i},
\end{align}
where $A_{eff}(E)$ is the detector effective area, $N_H(E)$ is the galactic column absorption and $\sigma_T$ is the Thompson cross-section. The summation limits implied here by `\textit{region}' are $\{r_o$~to~$r_{SH}\}$ for $r_n > r_{SH}$ and $\{r_{SH}$~to~$r_i\}$ for $r_n<r_{SH}$. Here, $i$ denotes the energy band used, and so can take any value from the set [$Lo$, $Int$ or $Hi$].\linebreak$k(r_n)$ denotes the spectral component originating from annulus $r_n$, so that $k(r_n)=S + R_S$ for $r_n > r_{SH}$, and $k(r_n)=H + R_H$ for $r_n < r_{SH}$. $L_{k(r_n)}^{i}$ therefore denotes the luminosity produced by spectral component $k$, as seen in energy band, $i$.

Since we have assumed only two Compton components, $L_{k(r_n)}^{i}$ here can take one of only six values depending on the energy band and radial zone. Explicitly, we have
\begin{equation}
L_{k(r_n)}^{i}=
\begin{cases}
L_{S+R_S}^{i} & \text{if}\ r_n > r_{SH}, \\
L_{H+R_H}^{i} & \text{if}\ r_n < r_{SH}, \\
\end{cases}
\label{eq:free_weights2}
\end{equation}
for each of the three bands. Usefully, the rms normalisation on the power spectra and cross spectra of equations~(\ref{eq:PSDprop2band})~and~(\ref{eq:Complex_Cross_Spectrum}) mean that in calculating timing statistics, we actually require only the \textit{ratio} of luminosities from each component in each band, $J^{i}$ = $L_{S+R_S}^{i}/L_{H+R_H}^{i}$. We can also remove the necessity for a spectral prior in determining $r_{SH}$ by simply allowing $f_H^S$ (formerly defined by the spectral fit throigh equation~\ref{r_SH}) to be free. This leaves us with four additional free parameters in the absence of an independent spectral fit: $f_H^S$, $J^{Lo}$, $J^{Int}$, $J^{Hi}$.

\begin{table}
	\begin{center}
	\begin{tabular}{cccc}
		\hline
		 & \textbf{\textit{FSW1}} & \textbf{\textit{FSW2}} & Fixed by \\
		 & & & spectral fit \textbf{\textit{2CFR}} \\
		\hline
		\vspace{+3pt}
		PSDs, Lags & Fig.~\ref{fig:FREE003} & - & Fig.~\ref{fig:2CFR003} \\
		\vspace{+5pt}
		$J^{Lo}$ & $0.31$ & $5.02^{\dagger}$ & $4.21$  \\
		\vspace{+5pt}
		$J^{Int}$ & $0.16$ & $0.26$  & $0.13$ \\
		\vspace{+5pt}
		$J^{Hi}$ & $0.11$ & $0.02$  & $0.02$ \\
		\vspace{+5pt}
		$f_H^S$ & $4.97$  & $1.27$ & $1.32$ \\ 
		\hline
	\end{tabular}
	\end{center}
	\caption[caption]{Fit results for the additional free parameters included when all constraints from the spectrum are released. $J^{Lo}$, $J^{Int}$ and $J^{Hi}$ denote the luminosity ratios between the \textit{soft} and \textit{hard} Compton components in the Low, Intermediate and High bands respectively, while $f_H^S$ denotes the ratio of bolometric luminosities in the \textit{soft} and \textit{hard} components (see equation~\ref{r_SH}). In the cases of fits \textbf{\textit{FSW1}} and \textbf{\textit{FSW2}}, $J^{Lo}$, $J^{Int}$, $J^{Hi}$ and $f_H^S$ are included as free parameters in the fits to the timing statistics. For \textbf{\textit{FSW1}}, we allow the new parameters to all be free. For \textbf{\textit{FSW2}}, we constrain $J^{Lo}$ to be greater than 4.2, as this is implied by the FR spectra (see Section~\ref{Fourier Resolved - 2 component}). The final column is included only for comparison, and shows the parameters which would be imposed by spectral fit \textbf{\textit{2CFR}}.}
	\label{tab:spectral_weightings}
\end{table}

\begin{figure}
	\includegraphics[width=\columnwidth]{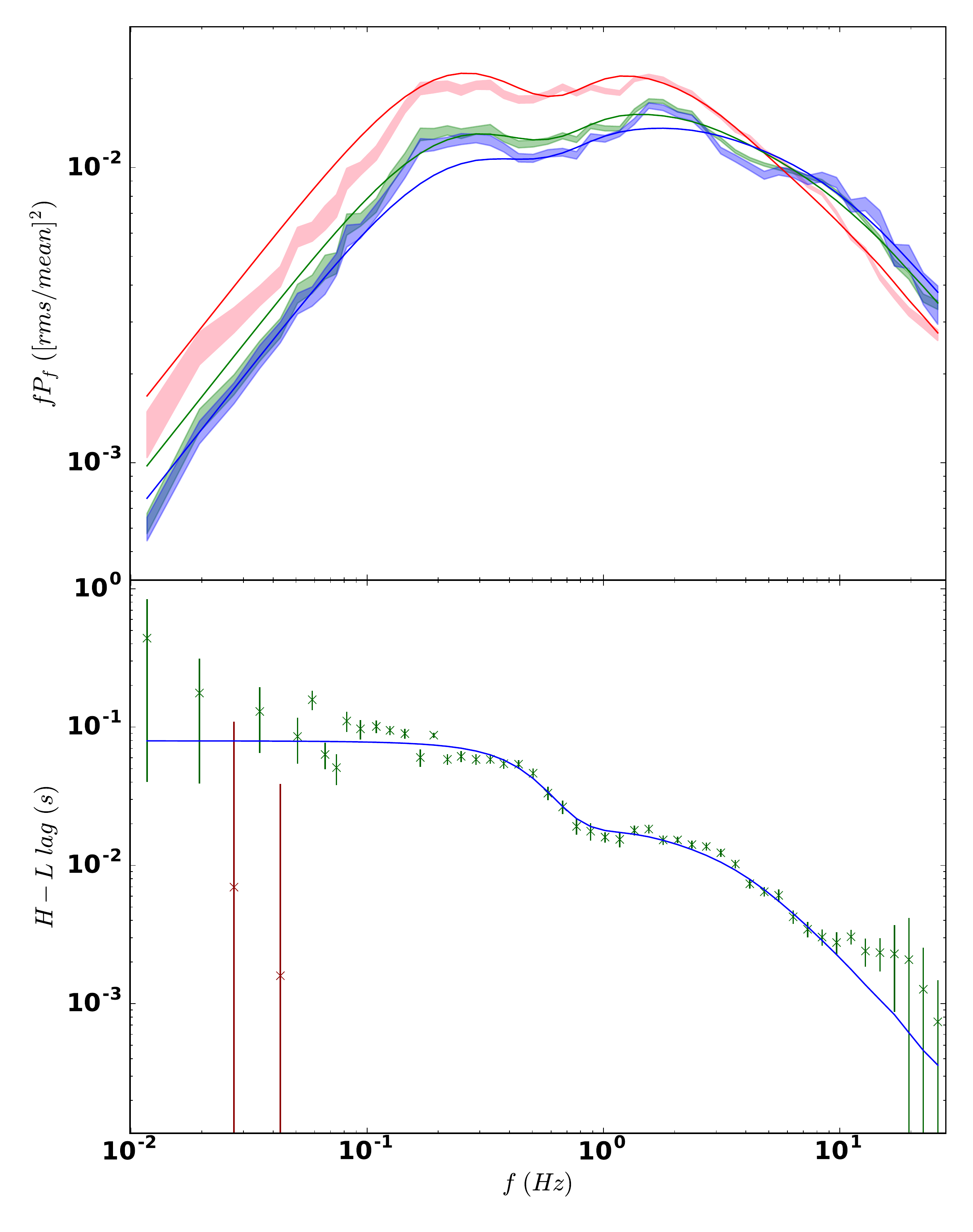}
	\caption{Timing fit with free spectral weights on all components, using the new free parameters of $J^{Lo}$, $J^{Int}$, $J^{Hi}$ and $f_H^S$. Top panel (a): High, Intermediate \& Low band PSDs. Colours as in Fig.~\ref{fig:MD17003}(a). Bottom panel (b): High-Low band time lags. Colours and symbols as in Fig.~\ref{fig:MD17003}(b).}
	\label{fig:FREE003}
\end{figure}

The best fit for our model free of spectral constraints is shown in Fig.~\ref{fig:FREE003}, with fit values for the new parameters shown in Table~\ref{tab:spectral_weightings}. We denote this `free spectral weight' fit as \textit{\textbf{FSW1}}. Now free from the constraints of the spectrum, this fit shows remarkable agreement with the data comapared to previous cases. The PSD hierarchies and structure are better reproduced than before, although the sharpness of the high-energy peaks seems to be slightly underpredicted, potentially indicating that the \textit{soft} component power should directly feed the \textit{hard} Component through seed photon variations, as well as mass fluctuations. We discuss this idea further in Section~\ref{sec:Conclusions}. Remarkably though, the lag structure is very well matched for the first time, showing that the lag structure can be reproduced by certain regions of the spectral-timing parameter space, but that constraints imposed by the spectral fit have so far prevented this structure from appearing. On examining the final fit values of $J^{Lo}$, $J^{Int}$ and $J^{Hi}$ in Table~\ref{tab:spectral_weightings}, it seems that the spectral picture required by the timing data is that of a \textit{soft} Compton component which decays monotonically with energy, providing less than $20\%$ contribution to the flux even in the Low energy band. This seems remarkably close to the behaviour of spectral fit \textit{\textbf{2C}}, and those of other fitting studies (e.g. \citealt{DS01}; \citealt{B17}) which have also indicated a \textit{hard} component which dominates at all energies above $\sim 1$~keV, with only a minute \textit{soft} compton contribution at low energies. The glaring problem with this spectral shape becomes clear however when looking at the required ratio of total fluxes in the \textit{soft} and \textit{hard} components, $f_H^S$. \textit{\textbf{FSW1}} requires $f_H^S\approx 5$, while models which give the required $J^{i}$ behaviour have \textit{soft} components with no more than $\sim 10\%$ of the luminosity of the hard component, e.g. in the case of \textit{\textbf{2C}} where $f_H^S \approx 0.06$. Not accounting for an unexpected sudden upturn in \textit{soft} Compton contribution below the PCA bandpass then, the values found for the $J^{i}$ and $f_H^S$ are in extreme tension. However, previous spectral fitting studies have also lacked a key tool which we can now use to further constrain the spectrum, extracted from the timing data itself. This tool is the Fourier-resolved (phase-resolved) spectrum.

\subsection{Constrained by the Fourier Resolved Spectra}
\label{Fourier Resolved - 2 component}

\begin{figure}
	\includegraphics[width=\columnwidth]{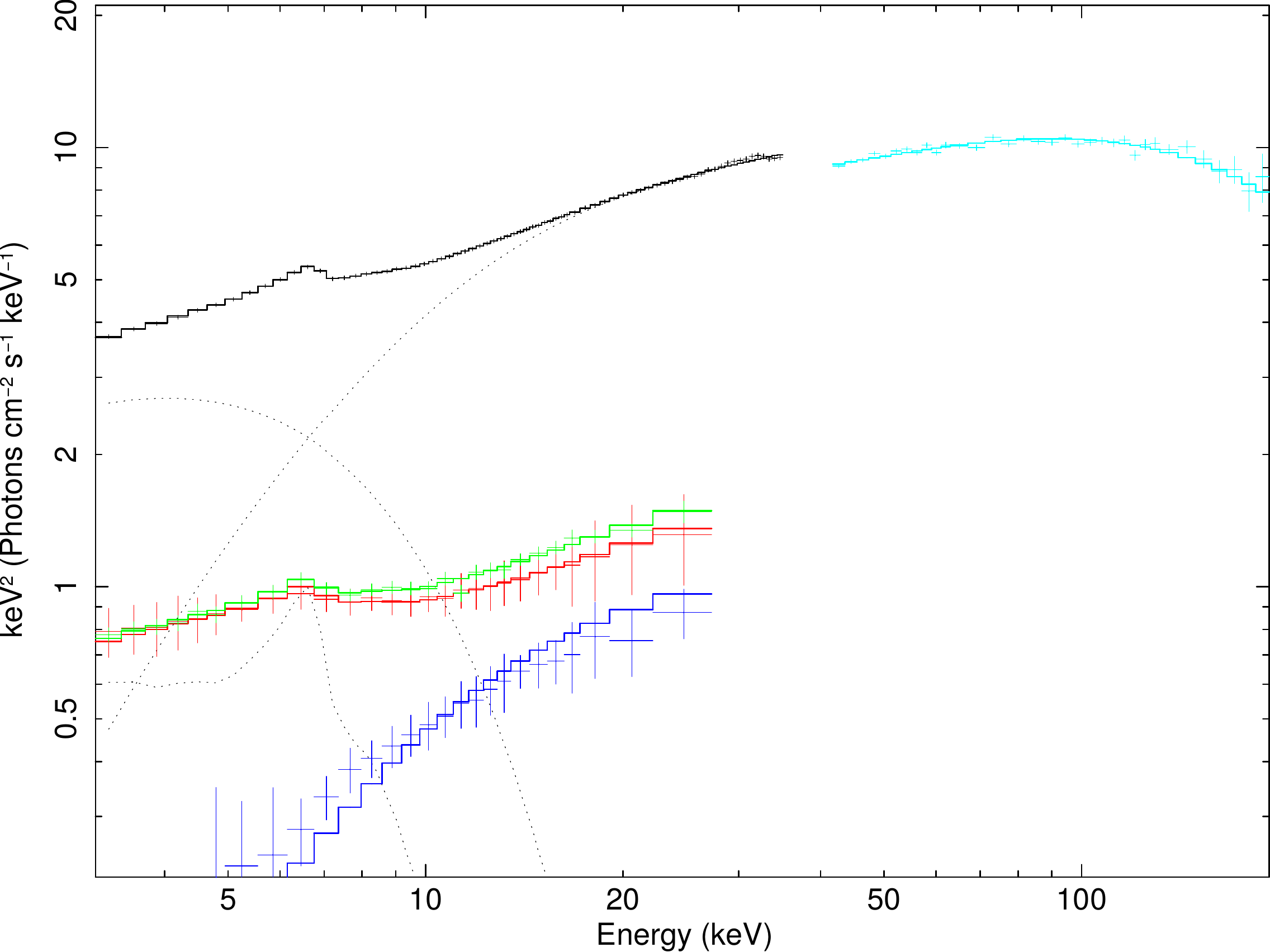}
	\caption{Two component simultaneous fit \textbf{\textit{2CFR}}, fit to the broadband spectrum from Obs. 1 and the Fourier-resolved components. Shown are the total energy spectrum in the PCA band (black crosses) and fit  (black solid line), with fit components (black dashed lines), and the total spectrum in the HEXTE band (cyan crosses) and fit (cyan solid line), with fit components (cyan dashed lines). We also show the FR spectrum for the slow variability (red crosses), and its fit (red solid line), the FR spectrum for the intermediate variability (green crosses), and its fit (green solid line), and the FR spectrum for the fast variability (blue crosses), and its fit (blue line). These FR components are fit using linear combinations of the \textit{soft}, \textit{mid} and \textit{hard} Compton components and their reflection, as detailed in the text.}
	\label{fig:2CFR_eeuf}
\end{figure}

\begin{figure}
	\includegraphics[width=\columnwidth]{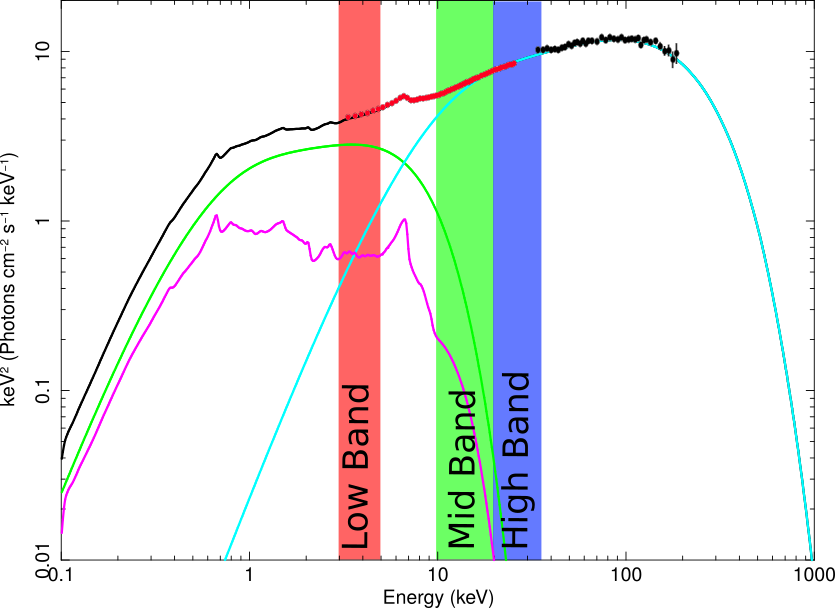}
	\caption{Broadband representation of spectral model \textbf{\textit{2CFR}}. Shown are the total energy spectrum (black), the \textit{hard} Compton component ($H(E)$, cyan), the \textit{soft} Compton component ($S(E)$, green), and the reflection component ($R(E)$, magenta). Filled circles show the PCA (red) and HEXTE (black) data. The red, green and blue bands denote the Low (3.13-4.98~keV), Intermediate (9.94-20.09~keV) and High (20.09-34.61~keV) energy ranges respectively.}
	\label{fig:2CFR}
\end{figure}

\begin{table*}
	\centering
	\begin{tabular}{lcccccccccccc}
		\hline
		Spectrum & $\Gamma_S$ & $kT_{e, S}$ & $n_S$ & $\Gamma_H$ & $kT_{e, H}$ & $kT_{seed, H}$ & $n_H$ & $R_{in}$ & $\left( \frac{\Omega}{2\pi}\right)_S$ & $\left( \frac{\Omega}{2\pi}\right)_H$ &log($x_i$) & $\chi^2/dof$  \\
		& & (keV)& & & (keV) & (keV) & $\times 10^{-2}$ & ($R_g$) & & & &   \\
		\hline
		\vspace{+3pt}
		Total & $1.85\pm 0.03$  & $1.7^{+0.1}_{-0.1}$ & $2.04^{+0.14}_{-0.07}$ & $1.69\pm 0.01$ & $105^{+13}_{-9}$ & $3.05 \pm 0.13$ & $2.3 \pm 0.2$ & $13^{+9}_{-4}$ & $-1.1 \pm 0.2$ & - & $3.3^{+0.2}_{-0.3}$ & $78.9/167$ \\
		\vspace{+3pt}
		Slow FR & "  & " & $0.43 \pm 0.03$ & " & " & " & $0.37^{+0.01}_{-0.02}$ & " &"&-& " &\\
		\vspace{+3pt}
		Int. FR & "  & " & $0.43 \pm 0.02$ & " & " & " & $0.41^{+0.03}_{-0.02}$ & " &"&-& " &\\
		\vspace{+3pt}
		Fast FR & n/a  & n/a & n/a & " & " & " & $0.262^{+0.007}_{-0.009}$ & n/a & n/a & n/a & n/a &\\ 
		\hline
	\end{tabular}
	\caption[caption]{Fitting parameters for the two component spectral model \textbf{\textit{2CFR}}, described by {\tt{tbnew_gas * (nthcomp + nthcomp + kdblur * xilconv * nthcomp)}}, fit to the total and FR spectra simultaneously. $n_m$ and $\left( \frac{\Omega}{2\pi}\right)_m$ denote the normalisation and reflection fractions on Compton component $m$. Uncertainties are quoted at the $1\sigma$ confidence level. Associated spectra are shown in Figs.~\ref{fig:2CFR_eeuf} \& \ref{fig:2CFR}.}
	\label{tab:2CFRspecparams}
\end{table*}

\begin{figure}
	\includegraphics[width=\columnwidth]{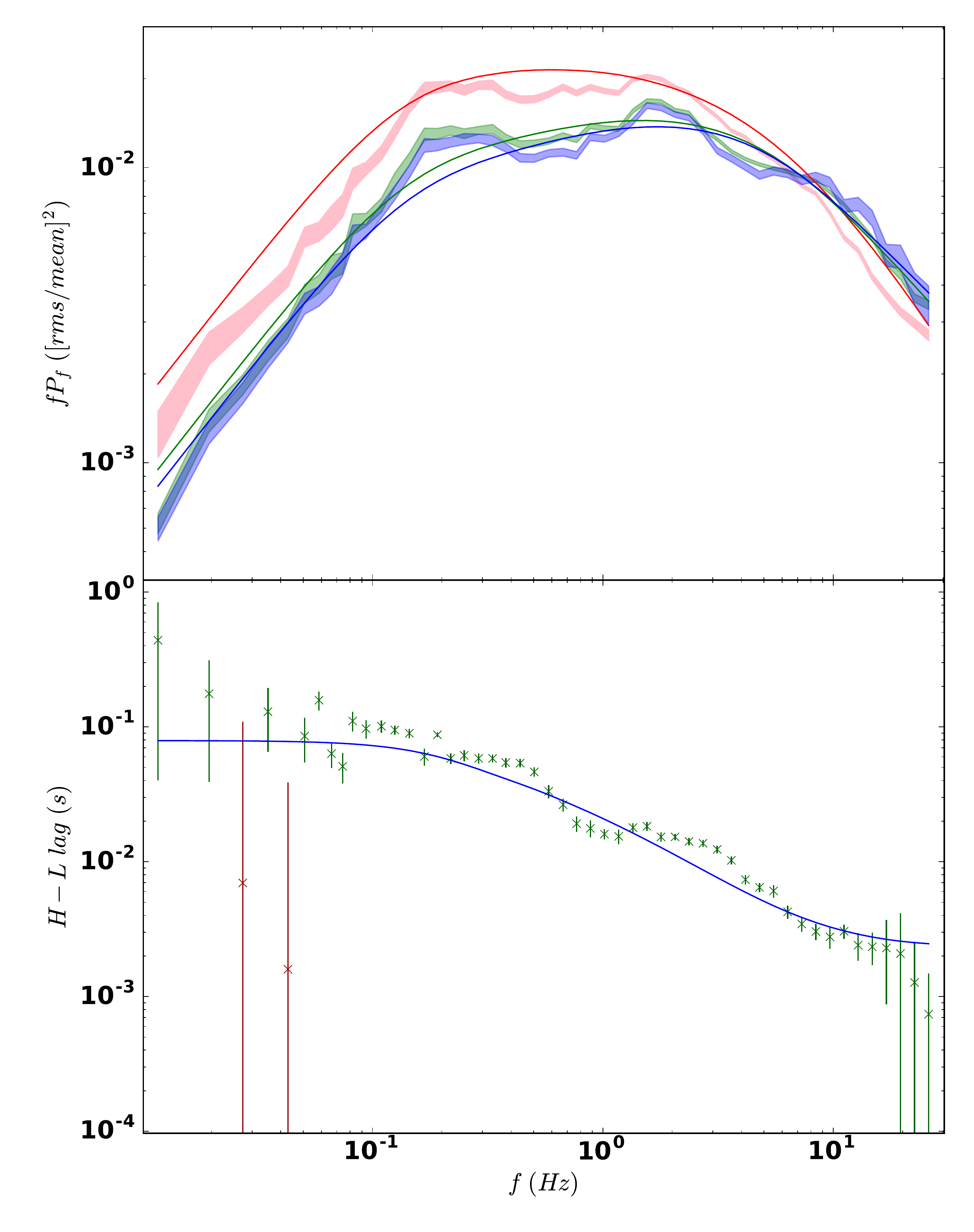}
	\caption{Timing fit using spectral model \textbf{\textit{2CFR}}. Top panel (a): High, Intermediate \& Low band PSDs. Colours as in Fig.~\ref{fig:MD17003}(a). Bottom panel (b): High-Low band time lags. Colours and symbols as in Fig.~\ref{fig:MD17003}(b).}
	\label{fig:2CFR003}
\end{figure}

\begin{figure}
	\includegraphics[width=\columnwidth]{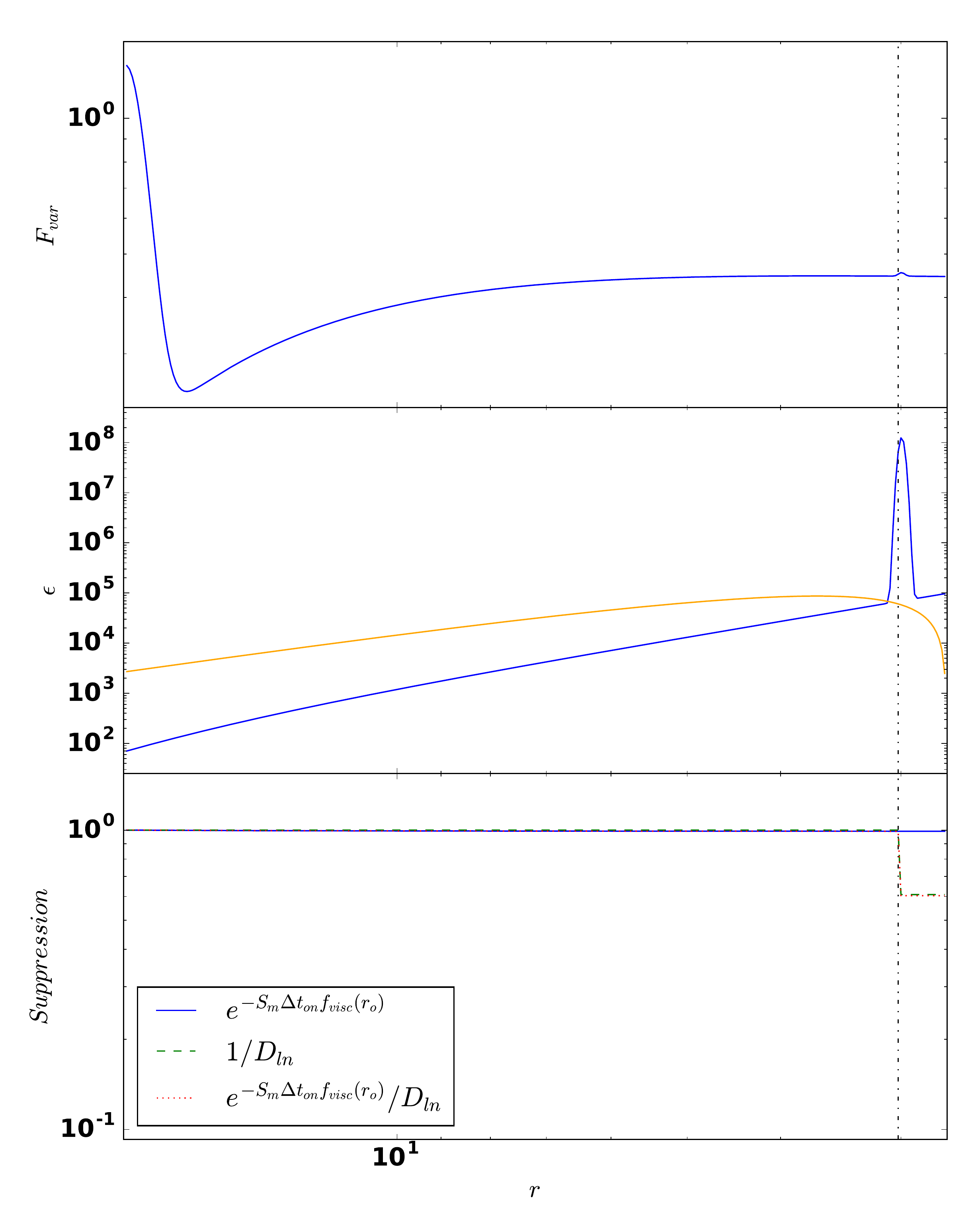}
	\caption{Fractional variability ($F_{var}$, top panel (a)), emissivity ($\epsilon$, middle panel (b)) and smoothing/damping (bottom panel (c)) profiles found for fit of Fig.~\ref{fig:2CFR003}. Colours and linestyles as in Fig.~\ref{fig:MD17003_fvar_emiss_suppression}.}
	\label{fig:2CFR003_fvar_emiss_suppression}
\end{figure}

To place further constraints on the spectral fit, we now draw on the Fourier-resolved (FR) spectra found using the technique of \cite{AD18}. These spectra are derived from the first, second and third humps in the energy-independent power spectrum of Obs. 1 (designated here as the `slow', `intermediate' and `fast' variability respectively). In Fig.~\ref{fig:2CFR_eeuf}, the slow, intermediate and fast FR spectra are shown as the red, green and blue crosses respectively. 

The most obvious fact on comparison of the FR spectra is that the fast component is highly distinct from the other two, with a much steeper slope down to its minimum resolved energy. Since this fast variability corresponds to where $f_{visc}(r)$ is highest, it must derive from the innermost, and therefore the hardest, region. The fast FR component therefore provides a strong constraint on the shape of the \textit{hard} Compton component, showing that its seed photon temperature, $kT_{seed,\,H}$, should be much higher than that of the \textit{soft}, which draws from the (shredding) disc at $kT_{seed,\,S} = 0.2$ keV. If the seed photons for the \textit{hard} component are coming from a source other than the disc (e.g. the \textit{soft} region), then the \textit{hard} region cannot have a line of sight to the disc. If the \textit{hard} region cannot intercept the photons from the disc, then the disc in turn must not intercept photons from the \textit{hard} region, and so there is no way that the \textit{hard} component can produce disc reflection. In the following spectral fits, we therefore switch off reflection from the \textit{hard} component [$R_H(E)$=0], in order to keep the spectral model self-consistent.

We now simultaneously fit to the broadband energy spectrum of Obs. 1 and its FR spectra, using 1\% systematic errors in the PCA bandpass to ensure consistency with \cite{AD18}. Unlike spectral fit \textbf{\textit{2C}}, we allow the \textit{hard} component seed photon temperature to be free, and untie the \textit{soft} and \textit{hard} electron temperatures. Since the slow variablity is produced in the outer regions, it will propagate through both the \textit{soft} and \textit{hard} Compton regions. The slow FR spectrum (red crosses with red line fit) is therefore fit with a linear combination of the \textit{soft} and \textit{hard} {\tt{nthcomp}} components, along with the reflection from the \textit{soft} component. We allow all normalisations of these components to be free, as it is unknown how much of the slow variability propagates into the \textit{hard} region. The intermediate FR spectrum (green crosses and solid line in Fig.~\ref{fig:2CFR_eeuf}), whose spectral shape is very similar to that of the slow, is similarly fit with a linear combination of the \textit{soft} and \textit{hard} {\tt{nthcomp}} components and \textit{soft} reflection, again with free normalisations. Finally the fast FR spectrum (blue crosses and solid line in Fig.~\ref{fig:2CFR_eeuf}) is fit only with the \textit{hard} {\tt{nthcomp}} component and no reflection, as this variability is produced and seen only in the \textit{hard} innermost region, and as mentioned has no direct line-of-sight to the disc. We denote this as spectral fit \textbf{\textit{2CFR}} (two component Fourier resolved), the parameters for which are shown in Table~\ref{tab:2CFRspecparams}. Despite the additional constraints from the phase-resolved spectra, untying the electron temperatures and the \textit{hard} Compton seed photon temperature has produced a significant improvement in the fit quality ($\chi^2_{\nu}= 78.9/167$) compared to spectral fit \textbf{\textit{2C}}. The overfitting in this case is due in part to the significant errors on the Fourier resolved spectra; discarding the FR spectra from this model we obtain ($\chi^2_{\nu}= 56.7/96$). While this is still overfit, this is often the case when applying multiple Compton components to data where the statistical errors underpredict the true uncertainty, and systematic errors must be relied upon (e.g. \citealt{I05}; \citealt{AD18}). However, an overfit in this case is acceptable since we know that additional spectral complexity is required by the timing properties, and that homogeneous Comptonisation models are chronically incapable of fitting such data (\citealt{I05}; \citealt{Y13}; \citealt{AD18}).

It is clear from the broadband representation of Fig.~\ref{fig:2CFR} that we now have a very different spectral picture from that of \textbf{\textit{2C}}, where the \textit{soft} Compton component is now dominant until the iron $K\alpha$ line, switching abruptly to \textit{hard} Compton dominance.

Using \textbf{\textit{2CFR}} gives the best fit combined spectral-timing shown in Fig.~\ref{fig:2CFR003}. The general shape of both the power spectra and the time lags are approximated. The double-broken power law behaviour is achieved with a switch in band dominance at $7-8$~Hz, as observed, while the lags decrease with frequency with the correct slope. However, the finer structure is not well modelled. The observed power spectra
show distinct humps whereas the model is much smoother, 
while the lags behave as a smooth power law function instead of displaying the clear steps seen in the data. These properties are due to the lack of structure present in the $F_{var}$ profile within $10$~$R_g$, where the variability is almost uniform save for a minute peak near $r_{SH}$. This profile was likely converged upon as a compromise between the approximate shape of the PSDs and lags, but it is clearly not a good representation of the timing data.

While spectral fit \textbf{\textit{2CFR}} does constrain the shape of the \textit{hard} Compton component, degeneracy still remains between the normalisation of this component and the shape of the \textit{soft} component (and its reflection). To exhaust our analysis of the two component class of spectral models, we would like to determine whether there exists \textit{any} spectral fit with two Compton components which satisfies the fast FR spectral shape, fits the energy spectrum, and provides a good fit to the timing properties. To do this, we can use the framework developed in Section~\ref{Weights_Free}, with one additional constraint.

In spectral fit \textbf{\textit{2CFR}}, the fastest FR component (blue lines and crosses in that figure) was well fit only with the \textit{hard} Compton component. The curvature of the \textit{hard} component is therefore well constrained until it peaks. In the case of that fit, the \textit{hard} component normalisation is maximal, since it contributes all of the photons at energies above $\sim25$~$keV$. This imposes a lower bound on the ratio of \textit{soft} to \textit{hard} component integrated flux in the Low energy band ($J^{Lo}$) of 4.2; for any alternative fit which replicates the fast FR spectrum, the \textit{hard} component normalisation may reduce, but it cannot become larger, and its shape cannot change.

We therefore apply the constraint of $J^{Lo}$ > 4.2 to our `free spectral weight' formalism of Section~\ref{Weights_Free}, to determine whether a better fit exists given only this requirement on the spectrum. We denote this fit as \textit{\textbf{FSW2}}, with fit parameters shown in Table~\ref{tab:spectral_weightings}. This yields a PSD/cross-spectral best fit which shows no improvement in fit quality over Fig.~\ref{fig:2CFR003} where spectral model \textbf{\textit{2CFR}} was used as a prior; for brevity, we therefore do not show this fit here. From this analysis it appears that no two component model exists which can reproduce the detailed structure in the timing properties simultaneously with satisfying the FR spectra. In particular the lag steps, which are obvious not only in these data, but also in the Astrosat/LAXPC observations of \cite{M17}, are entirely absent when using two Compton components. To find a model which can reproduce the timing features we therefore need to examine more complex spectral decompositions than the two component case, considering a more highly stratified flow than might be suggested by spectral fitting alone.

\section{Spectral Fit: Three Compton Components}
\label{Spectral Fit: Three Compton Components}

We now consider the decomposition of our energy spectrum into not two, but three distinct Comptonisation components, along with reflection from a truncated disc. This is not only motivated from the findings of Section~\ref{sec:2Comp}, but also from those of other spectral-timing analyses of the low/hard state. In particular, the phase-resolved analysis of \cite{Y13} showed that a highly variable, very soft component was necessary, in addition to two harder Compton components, to explain the spectral shape in the $0.7-300$~keV band for Cygnus X-1 data between 2005 and 2009.

\begin{figure}
	\includegraphics[width=\columnwidth]{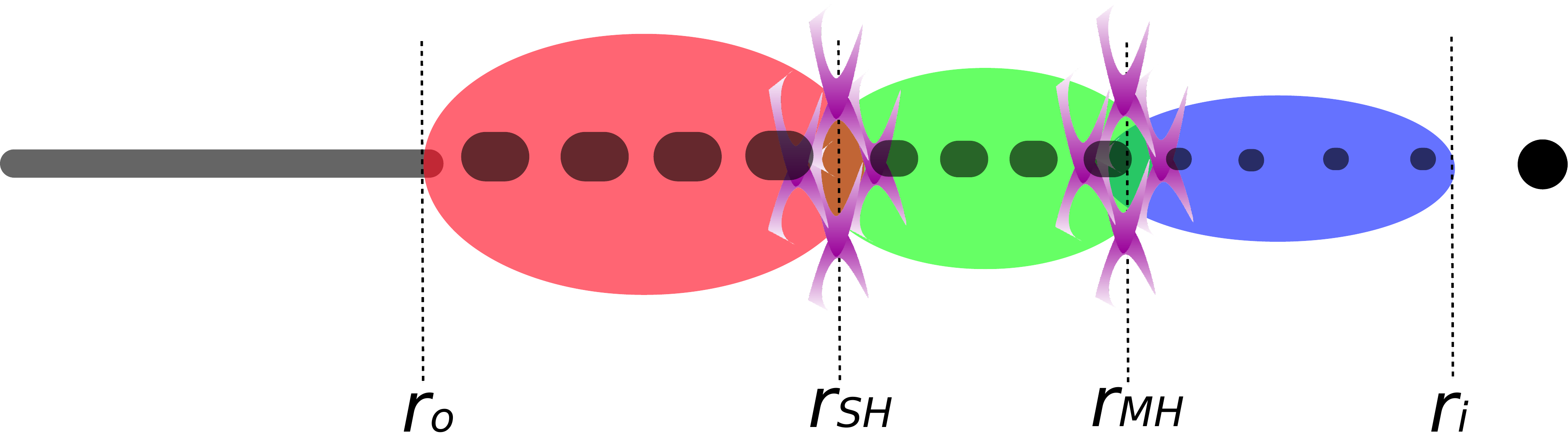}
	\caption{The physical geometry assumed within the flow in the three Compton component model. The dark grey region denotes the thermal, thin disc which does not vary on fast timescales. The red region denotes the fast-varying, spectrally \textit{soft} zone; the green region denotes the fast-varying, \textit{mid} spectral zone; the blue region denotes the fast-varying, spectrally \textit{hard} zone. Mass accretes down the flow from the disk truncation radius at $r_o$, through the \textit{soft}-\textit{mid} transition radius at $r_{SM}$ and then the \textit{mid}-\textit{hard} transition radius at $r_{MH}$, toward the inner flow radius at $r_i$. The purple knots in the flow denote a higher density of magnetic field lines at the spectral transition radius. The grey `blobs' in the flow represent thermal clumps, torn from the disc at the truncation radius, dissipating as they accrete. Clump dissipation, enhancement of the magnetic flux density near the transition radius, or a combination of the two phenomena would result in damping of fluctuations as they propagate from the \textit{soft} to the harder regions as required by the data, although our model does not distinguish between these mechanisms.}
	\label{fig:ThreeGeometry}
\end{figure}

\begin{figure}
	\includegraphics[width=\columnwidth]{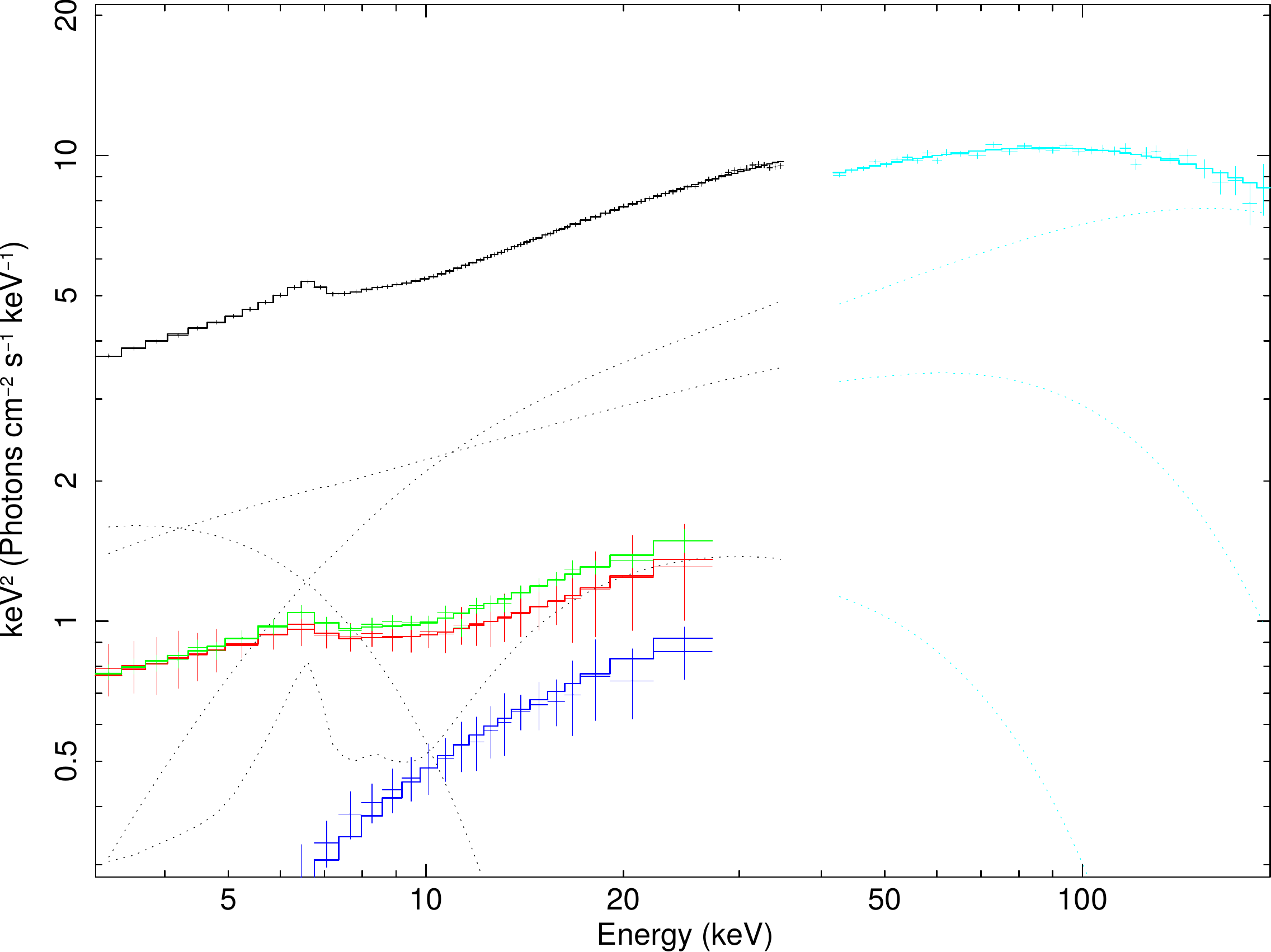}
	\caption{Unfolded three component spectral model \textbf{\textit{3CFR}}, fit to the broadband energy spectrum from Obs. 1 and the FR components. Shown are the total energy spectrum in the PCA band (black crosses) and fit  (black solid line), with fit components (black dashed lines), and the total spectrum in the HEXTE band (cyan crosses) and fit (cyan solid line), with fit components (cyan dashed lines). We also show the FR spectrum for the slow variability (red crosses), and its fit (red solid line), the FR spectrum for the intermediate variability (green crosses), and its fit (green solid line), and the FR spectrum for the fast variability (blue crosses), and its fit (blue line). These FR components are fit using linear combinations of the \textit{soft}, \textit{mid} and \textit{hard} Compton components and their reflection, as detailed in the text.}
	\label{fig:ThreeSpec1_eeuf}
\end{figure}

\begin{figure}
	\includegraphics[width=\columnwidth]{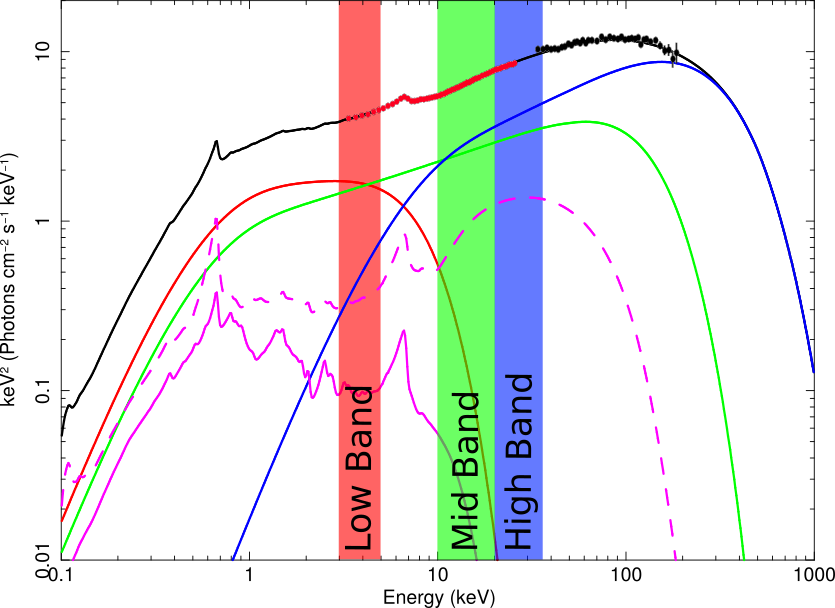}
	\caption{Broadband representation of three component spectral model \textbf{\textit{3CFR}}. Shown are the total energy spectrum (black), the \textit{hard} Compton component ($H(E)$, blue), the \textit{mid} Compton component ($M(E)$, green) the \textit{soft} Compton component ($S(E)$, red), and the reflection component ($R(E)$, magenta). Filled circles show the PCA (red) and HEXTE (black) data. The red, green and blue bands denote the Low (3.13-4.98~keV), Intermediate (9.94-20.09~keV) and High (20.09-34.61~keV) energy ranges respectively.}
	\label{fig:ThreeSpec1}
\end{figure}

\begin{table*}
	\centering
	\begin{tabular}{lccccccccccccccr}
		\hline
		Spectrum& $\Gamma_S$ & $kT_{e, S}$ & $n_S$ & $\Gamma_M$ & $kT_{e, M}$ & $n_M$ & $\Gamma_H$ & $kT_{e, H}$ & $kT_{seed, H}$ & $n_H$ & $R_{in}$ & $\left( \frac{\Omega}{2\pi}\right)_S$ & $\left( \frac{\Omega}{2\pi}\right)_M$ & log($x_i$) & $\chi^2/dof$  \tabularnewline
		& & (keV) & & & (keV)& & & (keV) & (keV) & $\times 10^{-2}$& ($R_g$) & & & & \tabularnewline
		\hline
		\vspace{+3pt}
		Total  & $1.91$  & $1.67$ & $1.36$ & $1.67$ & $36.1$ & $0.896$& $1.54$ & $166$ & $2.36$ &$1.75$& $12.9$ & $-0.539$ & $-0.382$ & $2.90$ & $60.1/158$ \tabularnewline
		\vspace{+3pt}
		Slow FR & "  & " & 0.387 & " & " & 0.117 & " & "& " & 0.335 & " &"& " & " &\tabularnewline
		\vspace{+3pt}
		Int. FR & "  & " & 0.254 & " & " & 0.224 & " & "& " & 0.160 & " &"& " & " &\tabularnewline
		\vspace{+3pt}
		Fast FR & n/a  & n/a & n/a & n/a & n/a & n/a & " & "& " & 0.399 & n/a & n/a & n/a & n/a & \tabularnewline 
		\hline
	\end{tabular}
	\caption[caption]{Fitting parameters for the three component spectral model \textbf{\textit{3CFR}}, described by {\tt{tbnew_gas * (nthcomp + nthcomp + nthcomp + kdblur * xilconv * (nthcomp + nthcomp))}}, fit to the total and FR spectra simultaneously. $n_m$ and $\left( \frac{\Omega}{2\pi}\right)_m$ denote the normalisation and reflection fractions on Compton component $m$. Associated spectra are shown in Figs.~\ref{fig:ThreeSpec1_eeuf} \& \ref{fig:ThreeSpec1}.}
	\label{tab:3specparams1}
\end{table*}

In our formalism we now include an additional Compton component between the \textit{soft} and \textit{hard} - the ``\textit{mid}" component, $M(E)$, with reflection $R_M(E)$. In the spectral timing model, the flow is therefore radially stratified into three distinct regions with \textit{soft}, \textit{mid} and \textit{hard} spectral shapes, illustrated in Fig.~\ref{fig:ThreeGeometry}. The time-averaged flux from each annulus is therefore
\begin{equation}
\label{F_r3}
\bar{F}(E, r_n)=
\begin{cases}
S(E) + R_S(E) & \text{if}\ r_n > r_{SM}, \\
M(E) + R_M(E) & \text{if}\ r_{MH} < r_n < r_{SM}, \\
H(E) & \text{if}\ r_n < r_{MH},
\end{cases}
\end{equation}
where we continue to neglect reflection from the \textit{hard} spectral component due to the fast FR spectral shape. Here $r_{SM}$ and $r_{MH}$ are the transition radii between the \textit{soft} and \textit{mid}, and \textit{mid} and \textit{hard} Comptonisation regions respectively, analytically derived from the prescribed emissivity such that the luminosity ratios between the Compton components matches that from the emissivity. These radii therefore satisfy the coupled equations
\vspace*{5 pt}
\begin{equation}
\begin{split}
\label{r_SM}
&\frac {\int_E S(E) dE}{\int_E M(E) dE} = \frac {\int_{r_o}^{r_{SM}} \epsilon(r)
	2\pi r dr }{\int_{r_{SM}}^{r_i} \epsilon(r)  2\pi r dr }, \\
&\frac {\int_E M(E) dE}{\int_E H(E) dE} = \frac {\int_{r_{SM}}^{r_{MH}} \epsilon(r)
	2\pi r dr }{\int_{r_{MH}}^{r_i} \epsilon(r)  2\pi r dr }.
\end{split}
\end{equation}

We now fit the time averaged SED, with 1\% systematic errors in the PCA bandpass, with three Comptonisation components described by {\tt{tbnew_gas * (nthcomp + nthcomp + nthcomp)}}, and the reflection of the \textit{soft} and \textit{mid} components, {\tt{tbnew_gas * (kdblur * xilconv * (nthcomp + nthcomp))}}. We simultaneously fit this model to both the total and FR spectra. In fitting we assume that the slow variability produced in the outer regions will be seen throughout the flow, and therefore in the \textit{soft}, \textit{mid} and \textit{hard} Compton components. The slow FR spectrum (red crosses and solid line in Fig.~\ref{fig:ThreeSpec1_eeuf}) is therefore fit with a linear combination of the \textit{soft}, \textit{mid} and \textit{hard} {\tt{nthcomp}} components, and the \textit{soft} and \textit{mid} reflection. As in Section~\ref{Fourier Resolved - 2 component}, we allow all normalisations of these components to be free. The intermediate variability (green crosses and solid line in Fig.~\ref{fig:ThreeSpec1_eeuf}) is also fit with a linear combination of the \textit{soft}, \textit{mid} and \textit{hard} {\tt{nthcomp}} components with reflection, as an improved fit was found when allowing this spectrum some contribution from all three Compton components. Finally the fast spectral component (blue crosses and solid line in Fig.~\ref{fig:ThreeSpec1_eeuf}) is satisfactorily fit only with the \textit{hard} {\tt{nthcomp}} component. We show this unfolded fit (denoted spectral model \textbf{\textit{3CFR}}, three component Fourier resolved) in Fig.~\ref{fig:ThreeSpec1_eeuf}, with the corresponding broadband shape shown in Fig.~\ref{fig:ThreeSpec1} and associated parameters shown in Table~\ref{tab:3specparams1}. This fit is somewhat akin to the spectral results of \cite{Y13}, with both fits featuring a very soft Compton component with a low-energy rollover in $S(E)$, in addition to the other two components, although our \textit{mid} component is significantly harder than the one found in their observation.

To allow the timing fit model to accomodate three spectral components, the damping parameter of equation~(\ref{2damp}) now takes the form
\begin{equation}
D_{ln} =
\begin{cases}
D_{SM} & \text{if } r_l \geq r_{SM} > r_n > r_{MH}, \\
D_{MH} & \text{if } r_{SM} > r_l > r_{MH} \geq r_n, \\
D_{SM}D_{MH} & \text{if } r_l \geq r_{SM}, r_{MH} \geq r_n, \\
1 & \text{otherwise.}
\end{cases}
\end{equation}
The viscous frequency must also be adapted to reflect the use of three spectral regions. This now follows
\begin{equation}
\label{eq:2visc_3comp}
f_{visc} = 
\begin{cases}
B_{S} r^{-m_S} f_{kep}(r) & \text{if } r\geq r_{SM} \\
B_{MH} r^{-m_{MH}} f_{kep}(r) & \text{if } r<r_{SM},
\end{cases}
\end{equation}
where $B_{MH}$ and $m_{MH}$ are the viscosity parameters which the \textit{mid} and \textit{hard} regions share; we do not include distinct timescales for each of these regions for model simplicity.

\begin{figure}
	\includegraphics[width=\columnwidth]{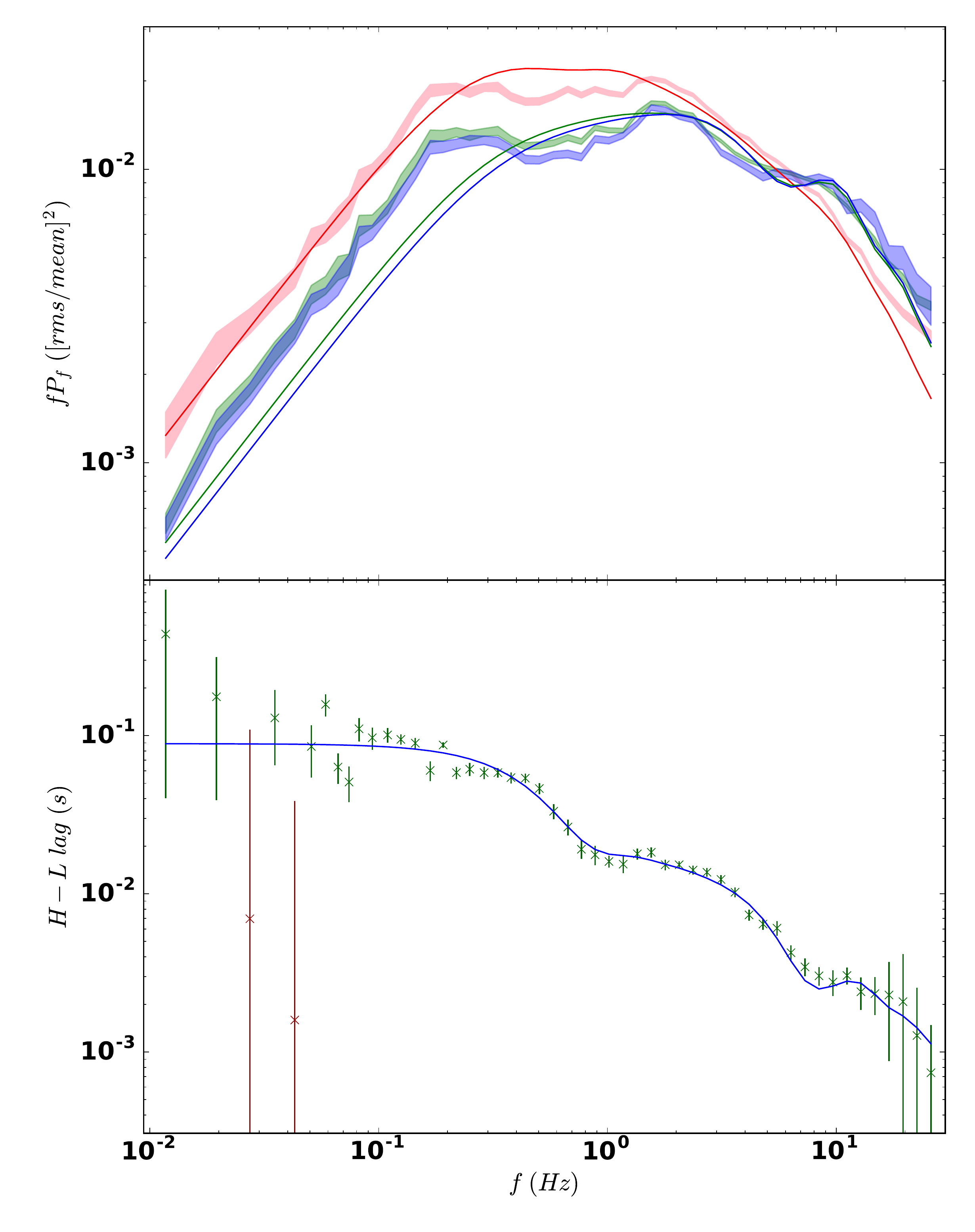}
	\caption{Timing fits using three component spectral fit \textbf{\textit{3CFR}}. Top panel (a): High, Intermediate \& Low band PSDs. Colours as in Fig.~\ref{fig:MD17003}(a). Bottom panel (b): High-Low band time lags. Colours and symbols as in Fig.~\ref{fig:MD17003}(b). A time-domain animated version of this fit, including the effect on the spectral components and the mass fluctuation behaviour through the flow can be found at: youtu.be/cGdqSAhxTVw}
	\label{fig:J2003}
\end{figure}

\begin{figure}
	\includegraphics[width=\columnwidth]{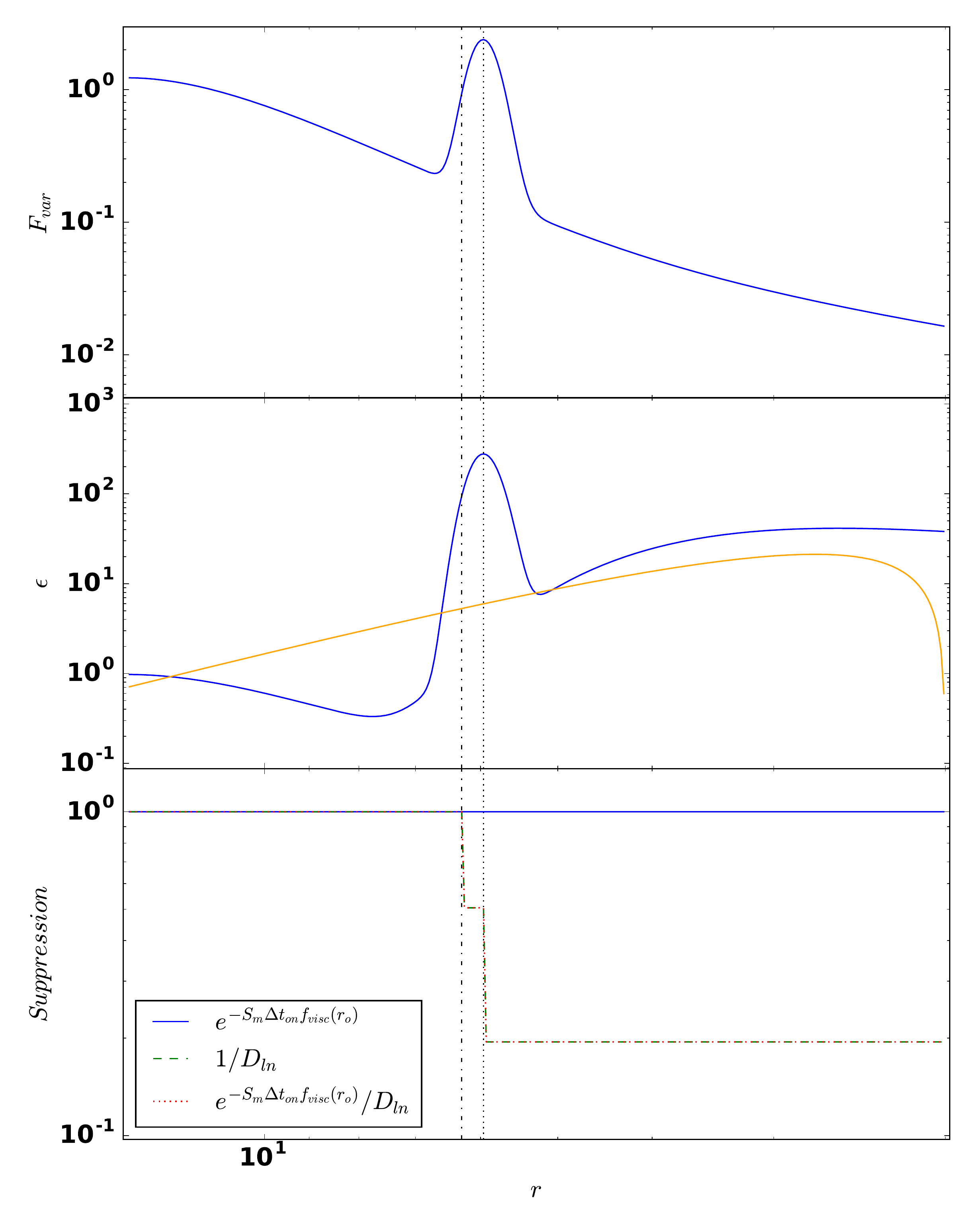}
	\caption{Fractional variability ($F_{var}$, top panel (a)), emissivity ($\epsilon$, middle panel (b)) and smoothing/damping (bottom panel (c)) profiles found for fit of Fig.~\ref{fig:J2003}. The black dot-dashed line denotes the \textit{soft}-\textit{mid} transition at $r_{SM}$. The black dotted line denotes the \textit{mid}-\textit{hard} transition at $r_{MH}$. Other colours and linestyles as in Fig.~\ref{fig:MD17003_fvar_emiss_suppression}.}
	\label{fig:J2003_fvar_emiss_suppression}
\end{figure}

Using spectral model \textbf{\textit{3CFR}}, we obtain a best fit to the timing properties shown in Fig.~\ref{fig:J2003}. The Intermediate and High band data are lacking in low-frequency power, likely due to the high degree of damping ($D_{SM}=1.98$ and $D_{MH}=2.59$). However, the Low energy PSD is a reasonable match to the data, with the broadband behaviour approximated, and most importantly we now find an excellent fit to the lag spectrum at all frequencies, exhibiting clear step features. The overall fit remains imperfect, but this agreement in the lags is particularly attractive. The inferred truncation radius of $13.8$~$R_g$ is also consistent with spectral fitting studies (\citealt{KDD14}, $16\pm 4$~$R_g$; \citealt{BZ16}, $13-20$~$R_g$). Examining the emission profile in Fig.~\ref{fig:J2003_fvar_emiss_suppression}(b), we find two bright ``rings" at $6$ and $3$~$R_g$. The turbulence profile of Fig.~\ref{fig:J2003_fvar_emiss_suppression}(a) also exhibits a significant peak at $6$~$R_g$, coincident with the spectral transition radii. This indicates a source of extreme variability and emission here, potentially giving rise to the switch in optical depth at this radius. Additional variability is also found at $r_o$, which would be attributable to disk-flow interaction. Enhanced emission (and to a lesser degree, enhanced variability) are similarly found in the timing fit from spectral model \textbf{\textit{2CFR}} (Figs.~\ref{fig:2CFR003}-\ref{fig:2CFR003_fvar_emiss_suppression}) indicating that the data require enhancement of the emission and/or turbulence at specific positions in the flow, independent of the true spectral complexity.

The qualitatively good fit we find in Fig.~\ref{fig:J2003} therefore strongly suggests that the fundamental model features required to approximate complex spectral-timing data such as these include at least: a non-constant $F_{var}$ profile, a bumpy emission profile corresponding to brighter annuli in the flow, at least three Compton components stratified with radius, and some form of damping of slow fluctuations as they propagate from softer to harder regions. Further complexities may be tested to fit the data in detail, but it appears that these features are required as a minimum to match the lags and PSDs.

Regarding the broken viscous timescale used in fitting, all timing model fits shown in this paper have also been run with a smooth viscous profile satisfying the $f_{lb}-f_{QPO}$ relation, i.e. $f_{visc} =0.03r^{-0.5}f_{kep}(r)$ for all $r$. Doing so produced only marginally worse fits in all cases (typical $\Delta \chi^2_{\nu} \sim 5\%$), although the lag structure in the case of Fig.~\ref{fig:J2003} is qualitatively much better reproduced with a broken viscous timescale. Since the inclusion of a smooth viscous timescale produces only a minor reduction in fit quality, we therefore do not regard a broken viscous timescale as a fundamental requirement of the model in the same manner as spectral stratification, $\epsilon(r)$/$F_{var}(r)$ enhancement, and fluctuation suppression.

While our model accommodates reasonable complexity in the Compton emission and turbulence, an endemic issue we have encountered in all fits has been the underprediction of higher energy, low-frequency power. However this could be rectified by the inclusion of spectral pivoting of the harder spectral components, driven by variations in the seed photon production from the softer region, as detailed in Section~\ref{Correlated Turbulence and Emissivity in a 2-Component Flow}. For spectral fit \textbf{\textit{2CFR}}, the position of the \textit{hard} seed photon temperature is quite close to the electron temperature of the \textit{soft} component ($kT_{seed,H}=3.05\pm 0.13$~keV compared to  $kT_{e,S}=1.6\pm 0.1$~keV), suggesting that at least some of the seed photons for the \textit{hard} component are from the \textit{soft} region. The variation in \textit{soft} region luminosity should therefore drive a change in the \textit{hard} component shape, and in turn should be an additional source of correlated variability between these components. In particular, our best fit case of Fig.~\ref{fig:J2003} approximates the High and Intermediate band spectra down to $\sim 1$~Hz, but cannot reproduce the low-frequency hump seen in these bands. Direct variability in the seed photons from the \textit{soft} component driving pivoting may help to correct this. If the \textit{hard} component pivot point is close to the High/Intermediate energy bands, pivoting would also result in  variability suppression in the High/Intermediate bands relative to the Low band, relaxing the requirement for variability damping as $\dot{m}$ fluctuations propagate from the \textit{soft} to harder regions. 

It is also worth noting that here we have shown the effect of using \textit{our} best fits to the total + FR spectra in the timing model. However, even with constraints from the FR spectra, systematic uncertainties on our data mean that the spectral fit itself is subject to degeneracies, and some other three component spectral fit may better represent reality in the accretion flow. Given that the three component model highly over-fits the spectral data alone ($\chi^2_{\nu}$ = 60.1/158), it is likely that a good spectral fit could still be obtained if the fitting procedure for the spectral and timing data were combined. This would provide better constraints on the accretion geometry by directly minimising across the model fits to both spectra and timing, but this is beyond the scope of this work.

\begin{table*}
	\begin{tabular}{lccc}
		\hline
		\vspace{+3pt}
		Spectral Model 			& \textbf{\textit{2C}} 			& \textbf{\textit{2CFR}} 			& \textbf{\textit{3CFR}} \\
		\hline
		\vspace{+3pt}
		PSDs, Lags 			& Fig.~\ref{fig:MD17003} 		& Fig.~\ref{fig:2CFR003} 		& Fig.~\ref{fig:J2003} \\
		\vspace{+3pt}
		$F_{var}(r)$, $\epsilon(r)$ 	& Fig.~\ref{fig:MD17003_fvar_emiss_suppression} 	& Fig.~\ref{fig:2CFR003_fvar_emiss_suppression} 	& Fig.~\ref{fig:J2003_fvar_emiss_suppression} \\
		\vspace{+3pt}
		$B_{S}$ 			& $0.03(F)$ 				& $0.03(F)$ 				& $0.03(F)$ \\
		\vspace{+3pt}
		$m_{S}$ 			& $0.5(F)$ 				& $0.5(F)$				& $0.5(F)$ \\
		\vspace{+3pt}
		$B_{H}$ 			& 1.36					& 0.269 				& n/a \\
		\vspace{+3pt}
		$m_{H}$ 			& 1.33					& 1.08	 				& n/a \\
		\vspace{+3pt}
		$B_{MH}$ 			& n/a 					& n/a 					& 0.0329 \\
		\vspace{+3pt}
		$m_{MH}$ 			& n/a 					& n/a 					& $3.88 \times 10^{-4}$ \\
		\vspace{+3pt}
		$r_o$ 				& 19.9					& 19.1 					& 13.8 \\
		\vspace{+3pt}
		$r_i$$^{\dagger}$ 		& 2.17					& 2.69 					& 2.00 \\
		\vspace{+3pt}
		$A_1$ 			& 2.15	 				& 1.13 					& 1.23 \\
		\vspace{+3pt}
		$A_2$ 			& 0.103	 				& 0.447 				& 2.22\\
		\vspace{+3pt}
		$A_3$ 			& 1.11					& $8.27 \times 10^{-3}$			& $5.01 \times 10^{-3}$ \\
		\vspace{+3pt}
		$r^{en}_1$ 			& $=r_o$ 				& $=r_o$ 				& $=r_o$ \\
		\vspace{+3pt}
		$r^{en}_2$ 			& 14.1 					& 3.64 					& 5.96 \\
		\vspace{+3pt}
		$r^{en}_3$ 			& 5.78 					& 3.00					& 2.58\\
		\vspace{+3pt}
		$\sigma^{en}_1$ 		& 2.56 					& 0.800					& 3.87 \\
		\vspace{+3pt}
		$\sigma^{en}_2$ 		& 1.79					& 11.6					& 0.213 \\
		\vspace{+3pt}
		$\sigma^{en}_3$ 		& 1.69	 				& 0.0216				& 1.40\\
		\vspace{+3pt}
		$Z_2$ 		& $2.48 \times 10^5$			& $2.45 \times 10^6$ 			& 280 \\
		\vspace{+3pt}
		$Z_3$ 		& 2400		 			& $4.53 \times 10^9$ 			& 41.7 \\
		\vspace{+3pt}
		$\gamma$ 			& 9.36 					& 3.25					& $9.31 \times 10^{-3}$ \\
		\vspace{+3pt}
		$S_m$$^\ddagger$ 		& $5.35 \times 10^{-3}$			& 0.0175				& $3.52 \times 10^{-6}$ \\
		\vspace{+3pt}
		${D}_{SH}$ 			& 3.90 					& 1.64 					& n/a \\
		\vspace{+3pt}
		${D}_{SM}$ 			& n/a 					& n/a 					& 1.98 \\
		\vspace{+3pt}
		${D}_{MH}$ 			& n/a 					& n/a 					& 2.59 \\
		\vspace{+3pt}
		$\chi^2/dof$ 			& 8481/289 				& 4875/289				& 4708/288 \\		
		\hline
		\multicolumn{4}{l}{\textsuperscript{$\dagger$}\footnotesize{A hard limit of $r_i \geq 2$ (near the typical ISCO)}}\\
		\multicolumn{4}{l}{\footnotesize{ is set for the inner radius of the hot flow.}}\\
		\multicolumn{4}{l}{\textsuperscript{$\ddagger$}\footnotesize{A hard limit of $S_m \geq 0$ is set so that fluctuations do}}\\
		\multicolumn{4}{l}{\footnotesize{not unphysically become narrower as they propagate.}}\\
	\end{tabular}
	\caption[caption]{Best-fitting parameter values for all timing models with a prior spectral fit shown in the body of this work. $(F)$ indicates that the parameter is fixed. We do not quote uncertainties due to the inherent parameter space degeneracies (discussed in Appendix~\ref{Appendix_Uncertainties}), and more importantly, the sensitivity to variation of the spectral model used as a prior.}
	\label{tab:params}
\end{table*}

\section{Conclusions}
\label{sec:Conclusions}
We expand on the spectral-timing model of MD18a describing the BHB low/hard state, in which fluctuations in the mass accretion rate propagate through a Comptonising, hot flow. This hot flow is stratified into spectrally distinct regions. We systematically explore the inclusion of more sophisticated profiles for the flow turbulence, emissivity,
damping/smoothing and spectral shape, and fit these models to some of the best data available which is from a bright low/hard state in Cygnus X-1. This data exemplifies the complex behaviour seen in many bright low/hard state BHBs, in which low energy bands dominate the power spectra at low frequencies, while high energy bands dominate at high frequencies (e.g. \citealt{WU09}; \citealt{G14}). These power spectra often contain distinct bumps, and the frequencies of this enhanced variability 
correspond to distinct  `steps' in the time lags as seen here (\citealt{M17}), as well as in other sources (e.g. in GX 339-4; \citealt{DM15}). 
Such data is very difficult to fit (R17a; \citealt{R17b}; MD18a), indicating a complex source geometry.
Lower luminosity low/hard states have 
smoother power spectra and time lags as a function of frequency, and can be fit with simpler models especially when the data only extend over a low energy bandpass where only the disc and outer parts of the hot flow
dominate (R17a; \citealt{R17b}). Here we use data from 3-35~keV, so the 
emission is dominated by the flow, and not by intrinsic disc emission. It is however an unfortunate fact that our selection criteria of a low/hard state source with the best signal-to-noise and the best telemetry RXTE PCA data resulted in an observation where the source structure is most complex.

The main results of this study can be summarised as follows:
\begin{enumerate}
\item {Multiple continuum components showing correlated (but not identical) variability are required to produce 
the different power spectra seen from different energy bands across the Comptonised emission, and the lags between them.}
\item {The data require that the amount of turbulence per radial decade is neither constant, nor a smooth power-law function of radius. Instead annuli of enhanced turbulence are present, separated by regions of low variability. These models therefore put less emphasis upon the MRI as the sole process driving the turbulence, as it is generally expected that the MRI will produce a uniform or at least smoothly varying amount of variability per radial decade. The enhanced turbulence may instead be due to disc-flow interaction at the truncation radius, non-axisymmetric tilt shocks, flow-jet interaction, and/or other even less well understood processes.}

\item {These regions of greater variability are also sometimes associated with additional emission. Unlike previous cases where the emission has been assumed to be a simple power-law function of radius (perhaps with an inner boundary condition), we find that the best fit cases require Compton-bright `rings' in the hot flow. Since the emission from these rings is enhanced, the photon flux variations from them are accentuated relative to the rest of the flow.}

\item {Having distinct regions of enhanced variability and emissivity is not by itself enough to make the distinct humps in the PSD. Propagation means that the power is cumulative at any frequency, so for the power to drop requires that not all the fluctuation power is transmitted to the next radius. 
The only exception to this is where the emission from two distinct regions is lagged such that interference can be important in supressing power (\citealt{V16}).}

\item{Damping of propagated fluctuations is also required so that lower energy bands show more low-frequency variability power than higher energy bands, which is why our models favour a damping term over the effects of interference.}
\end{enumerate}

We regard these as fundamental features required in any model 
which can reproduce the spectra, PSDs and time lags for the bright low/hard state. 
This forms the basis for a new paradigm for this state, whereby the Compton emission is dominated by bright, turbulent rings, i.e. a highly inhomogeneous accretion flow. Other sources in this state may be even more complex as there is often a strong QPO present for
higher inclinations.

Ultimately, even the use of spectral-timing data cannot break all the model degeneracies associated with the geometry and physical nature of the X-ray emission region without a better 
theoretical basis. The most fundamental issue is the lack of a clear prescription for the viscous frequency of the hot flow as a function of radius. This restricts our ability to convert from lag time to radius within the flow. The QPO-low frequency break relation can be used to determine the truncation radius between the disc and hot flow, but the viscous frequency within the rest of the hot flow is quite unconstrained. We are still a long way from numerical simulations which can include all of the potential physical components identified here - disc truncation and the torques from a misaligned flow and jet.

Nonetheless, there are much simpler PSDs seen in the lower luminosity low/hard states which presumably imply a simpler source structure, and there are now detections of reverberation lags from the central source illuminating the accretion disc which give independent size scale information in GX339-4 (\citealt{DMP16}; \citealt{DM17}). We will extend our method to fit these data in future work. 

\section*{Acknowledgements}
RDM and CD sincerely thank Magnus Axelsson for assistance with the Fourier-resolved spectral analysis and extensive discussions on the accretion flow behaviour. RDM acknowledges the support of a Science and Technology Facilities Council (STFC) studentship through grant ST/N50404X/1. CD acknowledges the STFC through grant ST/P000541/1 for support. This work used the DiRAC Data Centric system at Durham University, operated by the Institute for Computational Cosmology on behalf of the STFC DiRAC HPC Facility (www.dirac.ac.uk). This equipment was funded by BIS National E-infrastructure capital grant ST/K00042X/1, STFC capital grant ST/H008519/1, and STFC DiRAC Operations grant ST/K003267/1 and Durham University. DiRAC is part of the National E-Infrastructure. This research has made use of data obtained through the High Energy Astrophysics Science Archive Research Center Online Service, provided by the NASA/Goddard Space Flight Center.

\section*{Supporting Materials}
A time-domain animated version of the timing fit using spectral model \textbf{\textit{3CFR}} can be found at: youtu.be/cGdqSAhxTVw \\This video includes the mass accretion rate fluctuation behaviour through the flow, the effect on the spectral component variation, and the output time-domain light curves for a handy illustration of how the spectra and propagating fluctuations relate.

%%%%%%%%%%%%%%%%%%%%%%%%%%%%%%%%%%%%%%%%%%%%%%%%%%

%%%%%%%%%%%%%%%%%%%% REFERENCES %%%%%%%%%%%%%%%%%%

%%%%%%%%%%%%%%%%% APPENDICES %%%%%%%%%%%%%%%%%%%%%
\appendix

\section{Timing Formalism for Two Spectral Components}
\label{Timing Formalism}

IvdK13 show that in the absence of damping, the
propagated mass accretion rate curve at a given annulus can be written as:
\begin{equation}
\label{eq:mdot_propagation}
\dot{M}(r_n, t) = \prod_{l=1}^n \dot{M}(r_l, t - \Delta t_{ln}),
\end{equation}
where we denote capital $\dot{M}(r_n, t)$ as the \textit{propagated} mass accretion rate at radius $r_n$, distinct from the generated variability at $r_n$, $\dot{m}(r_n, t)$. This of course implies that $\dot{M}(r_o, t) = \dot{m}(r_o, t)$.

However recent work (R17a, \citealt{MIvdK17}) has highlighted that equation~(\ref{eq:mdot_propagation}) is only a specific case of the Green's function for the accretion of an annular unit mass from radius $r_l$ to $r_n$, $G(r_l, r_n, t)$. This function describes how a delta function perturbation initialised at $r_{l}$ spreads and propagates toward the compact object, to be observed at some inner annulus $r_n$. The general case of the generator functions is therefore given
by
\begin{equation}
\dot{M}(r_n, t) = \prod_{l=1}^n G(r_l, r_n, t) \circledast \dot{M}(r_l, t),
\end{equation}
where $\circledast$ denotes a convolution. In the case where $G(r_l, r_n, t)$ $=$ $\delta(t- t_{ln})$, we recover equation~(\ref{eq:mdot_propagation}) and the power spectrum of mass accretion rate fluctuations becomes the standard case of IvdK13,
\begin{equation}
\label{eq:Mdot_propagation}
|\tilde{\dot{M}}(r_n, f)|^2 = |\tilde{\dot{m}}(r_n, f)|^2 \circledast |e^{2\pi i \Delta t_{(n-1)n} f} \tilde{\dot{M}}(r_{n-1}, f)|^2.
\end{equation}

In reality, an infinitely narrow annulus propagating according to equation~(\ref{eq:Mdot_propagation}) would be unphysical, as it would simply move toward the black hole without dispersing radially. Instead, as the surface density fluctuations propagate, they are also expected to smooth out. For a radiatively inefficient accretion flow, the relation between the smoothing timescale and the local viscous timescale is poorly understood, so we follow R17a in parameterising the Fourier transform of this part of the Green's function as an exponential decay with frequency and lag time, scaled  by the smoothing parameter, $S_m$. This prescription has the property of smoothing out older and/or shorter timescale fluctuations first, as we would expect a physically realistic accretion flow to do.

In MD18a we also showed that the Low-band dominance of the PSDs at low frequencies demands that the seed photon variability from $r~>~r_{SH}$ be suppressed on propagation into the inner regions. Such seed photon suppression may arise from clump evaporation as thermal packets torn from the disc dissipate as they are accreted. We again model this generically by only including a fraction $D_{SH}^{-1}$ of the low frequency noise propagated from the outer regions in the time series of the inner regions. The effect of this damping is distinct from the diffusive smoothing process, in that it affects all fluctuations equally irrespective of frequency or of lifetime within the flow. This prescription is only the  simplest possible way to include damping, where the supression of amplitude takes place at a specific radius. It could instead be  a continuous function of radius, but since the final band-dependent light curves are simply weighted sums of the \textit{average} mass accretion rates in each spectral region, the discrete and continuous suppression cases are interchangeable in the model. The final Green's function therefore has a Fourier transform of
\begin{equation}
\label{eq:GreensFunction}
\tilde{G}(r_l, r_n, f) =
\frac{1}{D_{ln}} e^{2\pi i \Delta t_{ln} f} e^{-S_m \Delta t_{ln} f}.
\end{equation}

The propagated PSDs are now described by
\begin{equation}
\left|\tilde{\dot{M}}(r_n, f)\right|^2 =
\left|\tilde{\dot{m}}(r_n, f)\right|^2 \circledast \left| \tilde{G}(r_{n-1}, r_n, f) \tilde{\dot{M}}(r_{n-1}, f)\right|^2,
\end{equation}
which shows that the mass accretion rate in each annulus is simply a sequential convolution of all those preceeding it (rescaled by their Green's functions).
The modified equation in terms of the input generator power spectra is therefore
\begin{equation}
|\tilde{\dot{M}}(r_n, f)|^2 = \coprod^n_{l=1} \left|e^{\,-S_m \Delta t_{ln} f} \frac{\tilde{\dot{m}}(r_l,f)}{D_{ln}}\right|^2,
\end{equation}
where the coproduct symbol denotes sequential convolutions.

Our mass accretion rates are converted to counts in a given energy band, $i$, using
the emissivity prescription and SED decomposition described in Section
\ref{Correlated Turbulence and Emissivity in a 2-Component Flow}. This effectively weights the propagated mass accretion
rate from each annulus by a factor, $w_n^{\,i}$, given by
\begin{equation}
\label{eq:weights}
w_{n}^{\,i} = \frac{\epsilon(r_n) r_n dr_n}{\sum\limits_{region} {\epsilon(r_n) r_n dr_n}}\int\displaylimits_{E = E_{i}^{min}}^{E_{i}^{max}} \bar{F}(E, r_n)A_{eff}(E)e^{-N_H(E)\sigma_T} dE,
\end{equation}
where $A_{eff}(E)$ is the detector effective area, $N_H(E)$ is the galactic column absorption and $\sigma_T$ is the Thompson cross-section. The count rate for that band can then be written
\begin{equation}
C_{i}(t) = \sum_{n = 1} ^{N} w_n^{\,i} \dot{M}(r_n,t).
\end{equation}
Since the mean count rate of $\dot{M}(r_n, t)$ is normalised to $\dot{M}_0$, the mean count rate in a given energy band is then
\begin{equation}
\mu_C = \sum_{n = 1}^{N} \dot{M}_0 w_n^{i}.
\end{equation}
Dropping the superscript on $w_n^{i}$, the rms-normalised power spectrum of the variability in this energy band is then
\begin{equation}
\begin{aligned}
P_{i}(f) &= \frac{2dt^2}{\mu_{C}^2T}|\tilde{C}_{i}(f)|^2\\
&= \frac{2dt^2}{\mu_{C}^2T}\sum_{l,\,n=1}^{N} w_n w_l \tilde{\dot{M}}(r_l,f)^*\tilde{\dot{M}}(r_n,f).
\end{aligned}
\end{equation}

Since our Green's function has the property \linebreak $G(r_l, r_n, t)=G(r_l, r_k, t)\circledast~G(r_k, r_n, t)$, or equivalently $\tilde{G}(r_l, r_n, f)=\tilde{G}(r_l, r_k, f)~\tilde{G}(r_k, r_n, f)$, we can use the same arguments as the case of the Green's function of IvdK13 to show that, in the case of unity mean mass accretion rate at each annulus, the cross-spectrum between annuli can be expressed,
\begin{equation}\label{key}
\tilde{\dot{M}}(r_l,f)^*\tilde{\dot{M}}(r_n,f) = \frac{e^{2 \pi i \Delta t_{ln}}e^{\,-S_m \Delta t_{ln} f}}{D_{ln}}  \left|\tilde{\dot{M}}(r_l,f)\right|^2.
\end{equation}
The band-dependent power spectrum therefore becomes
\begin{equation}
\label{eq:PSDprop2band}
\begin{aligned}
P_{i}(f) = &\frac{2dt^2}{\mu_{C}^2T}\sum_{n=1}^{N}\left[w_n^2 |\tilde{\dot{M}}(r_n,f)|^2\right.\vphantom{...}\\ 
&+ 2\sum_{l=1}^{n-1}\left.\vphantom{..} w_l w_n \text{cos}(2 \pi \Delta t_{ln} f)\frac{|\tilde{\dot{M}}(r_l,f)|^2}{e^{\,S_m \Delta t_{ln} f} D_{ln}}\right]. 
\end{aligned}
\end{equation}

We can also compare the timing data in different energy bands
\begin{equation}
\label{eq:Complex_Cross_Spectrum}
\Gamma_{LH}(f) = \frac{2 dt^2}{\mu_L \mu_H T} \tilde{C}_{L}(f) ^* \tilde{C}_{H}(f),
\end{equation}
which yields a result for the cross-spectrum, $\Gamma_{LH}(f)$, analogous to that of equation~(\ref{eq:PSDprop2band}),
\begin{equation}
\label{eq:}
\begin{aligned}
\Gamma_{LH}(f)=&\frac{2dt^2}{\mu_{L}\mu_{H}T}\sum_{n=1}^{N}\left[w_n^{\,L}w_n^{\,H} |\tilde{\dot{M}}(r_n,f)|^2 \right.\vphantom{}\\
&+ \sum_{l=1}^{n-1}\left.\vphantom{} (w_l^{\,L}w_n^{\,H}e^{2\pi i \Delta t_{ln}f} +  w_l^{\,H} w_n^{\,L} e^{-2\pi i \Delta t_{ln}f})\frac{|\tilde{\dot{M}}(r_l,f)|^2}{e^{\,S_m \Delta t_{ln} f}D_{ln}}\right].
\end{aligned}
\end{equation}

Phase/time lags can then be obtained by splitting this cross spectrum into its real and imaginary parts,
\begin{equation}
\label{eq:CrossRe}
\begin{aligned}
\mathfrak{Re}[\Gamma_{LH}(f)]=&\frac{2dt^2}{\mu_{L}\mu_{H}T}\sum_{n=1}^{N}\left[w_n^{\,L} w_n^{\,H}  |\tilde{\dot{M}}(r_n,f)|^2\right.\vphantom{}\\ &+ \sum_{l=1}^{n-1} \left.\vphantom{} \text{cos}(2\pi \Delta t_{ln}f)(w_l^{\,L}w_n^{\,H} + w_l^{\,H} w_n^{\,L} )\frac{|\tilde{\dot{M}}(r_l,f)|^2}{e^{\,S_m \Delta t_{ln} f}D_{ln}} \right],
\end{aligned}
\vspace*{-10pt}
\end{equation}
\begin{equation}
\label{eq:CrossIm}
\begin{aligned}
\mathfrak{Im}[\Gamma_{LH}(f)] =\frac{2dt^2}{\mu_{L}\mu_{H}T}\sum_{n=1}^{N}\sum_{l=1}^{n-1}& \left[(w_l^{\,L}w_n^{\,H} - w_l^{\,H} w_n^{\,L} )\right.\vphantom{}\\
&\left.\vphantom{}\times \text{sin}(2\pi \Delta t_{ln}f)\frac{|\tilde{\dot{M}}(r_l,f)|^2}{e^{\,S_m \Delta t_{ln} f}D_{ln}}\right].
\end{aligned}
\end{equation}

From these complex components, the time lag is then extracted as

\begin{equation}
\label{eq:tau_analytic}
tan(2\pi f \tau_{LH})=\frac{\mathfrak{Im}[\Gamma_{LH}(f)]}{\mathfrak{Re}[\Gamma_{LH}(f)]},
\end{equation}
in the same way as the observed timing statistic, which allows direct comparison to the data.

\section{Spectral Timing Results for MD18a Model}
\label{MD18a results}

\begin{figure}
	\includegraphics[width=\columnwidth]{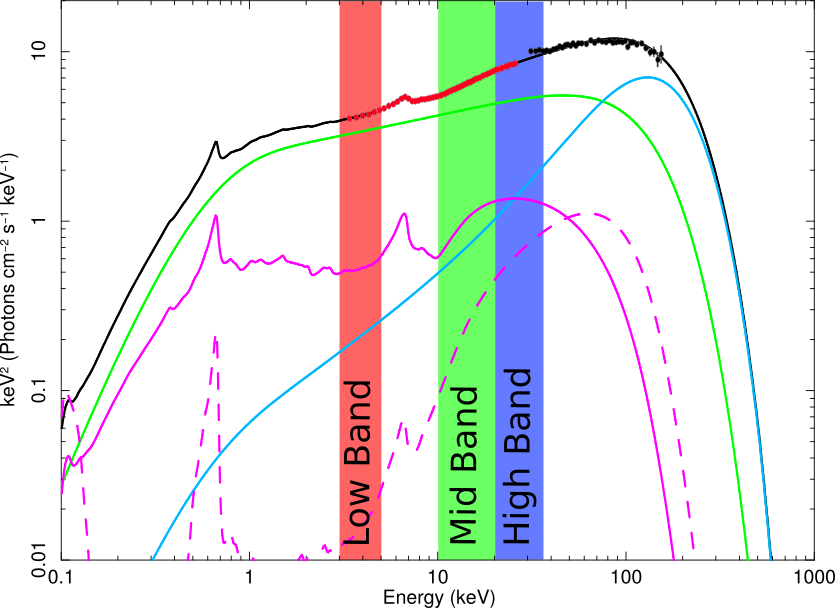}
	\caption{Two component spectral decomposition of Obs. 1 from MD18a. While the other fits in this paper use 1\% systematic errors in the PCA bandpass on the data, this fit uses 0.5\% systematics on the model at all energies. Lines show the total energy spectrum (black solid), the \textit{hard} Compton component ($H(E)$, cyan solid), the \textit{soft} Compton component ($S(E)$, green solid), the truncated disc reflection from the \textit{hard} component ($R_H(E)$, magenta dashed), and the reflection from the \textit{soft} component ($R_S(E)$, magenta solid). Filled circles show the PCA (red) and HEXTE (black) data. The red, green and blue bands denote the Low (3.13-4.98~keV), Intermediate (9.94-20.09~keV) and High (20.09-34.61~keV) energy ranges respectively. Systematic errors on model and data have been updated leading to very different spectral shape from that of MD18a which fit the same data with the same model.}
	\label{fig:MD17_old}
\end{figure}

\begin{figure}
	\includegraphics[width=\columnwidth]{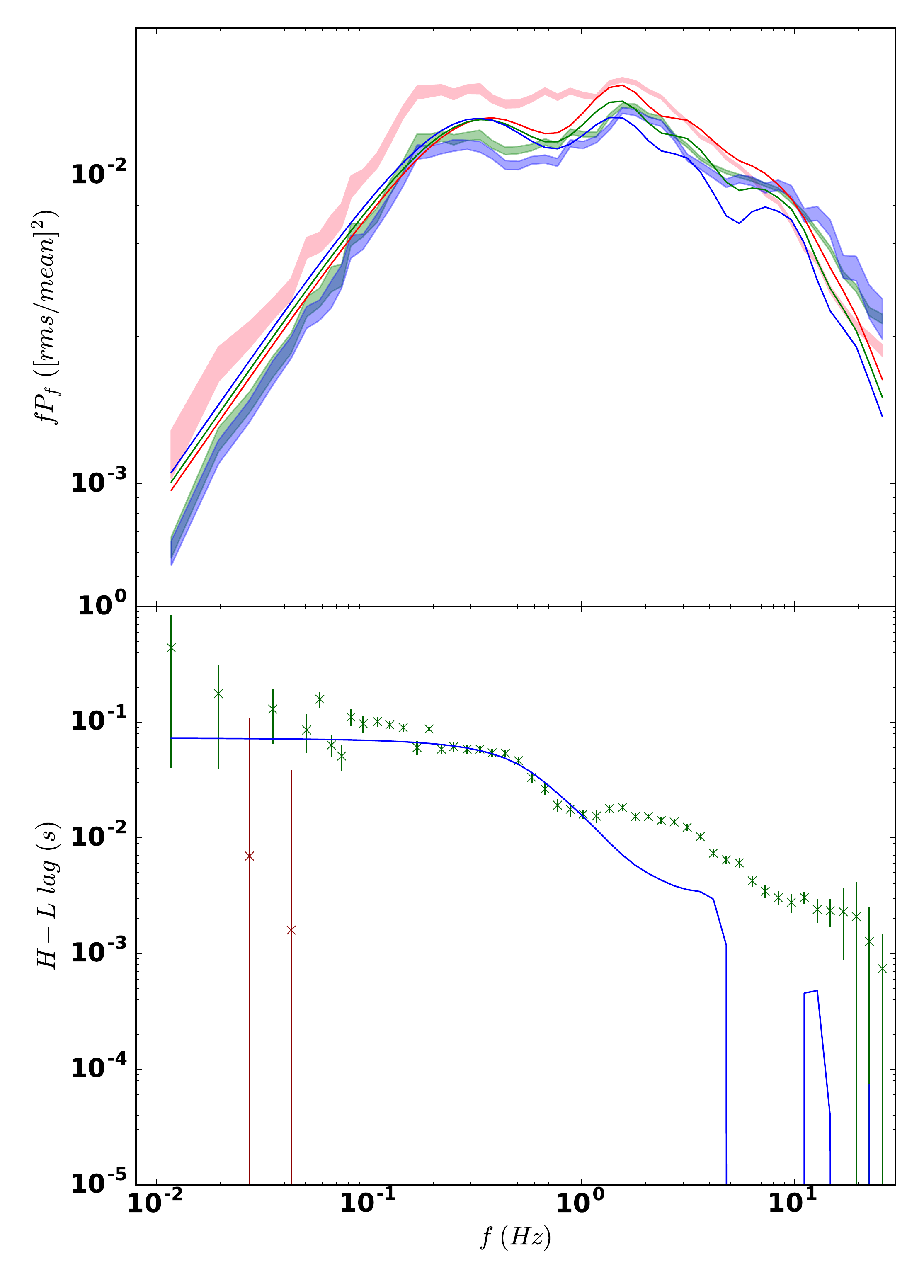}
	\caption{Timing fit using the spectral model of MD18a. Top panel (a): High, Intermediate \& Low band PSDs. The shaded regions are the 1$\sigma$ error regions of the Low (pink), Intermediate (green) and High (blue) energy bands from the data. The solid lines show the Low (red), Intermediate (green) and High (blue) energy model outputs. Bottom panel (b): Crosses denote the time lags between the High and Low bands for the data. Green crosses indicate the High band lagging the Low band. Red crosses indicate the Low band lagging the High band. The blue solid line denotes the model output.}
	\label{fig:MD17002}
\end{figure}

\begin{figure}
	\includegraphics[width=\columnwidth]{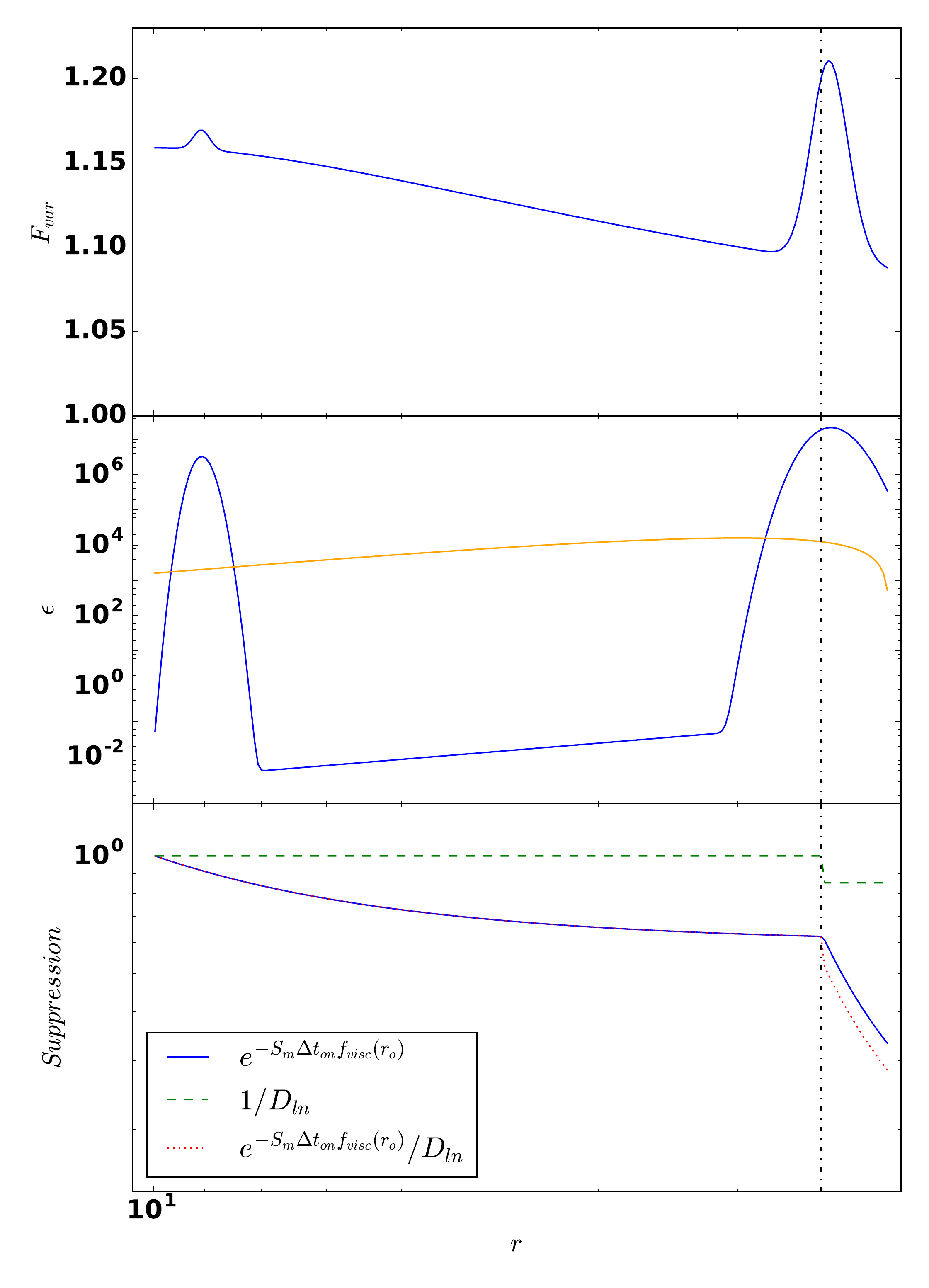}
	\caption{Top panel (a): Fractional variability ($F_{var}$) profile found for fit of Fig.~\ref{fig:MD17002}. Colours and linestyles as in Fig.~\ref{fig:Generic_Fvar_em_suppression}(a). Middle panel (b): Emissivity ($\epsilon$) profile found for fit of Fig.~\ref{fig:MD17002}. Orange solid line denotes Novikov-Thorne-type $\epsilon (r) \propto r^{-3} \left(1-\sqrt{r_i/r}\right)$ profile for comparison. Other colours and linestyles as in Fig.~\ref{fig:Generic_Fvar_em_suppression}(b). Bottom panel (c): Smoothing/damping profile found for fit of Fig.~\ref{fig:MD17002}. Colours and linestyles as in Fig.~\ref{fig:Generic_Fvar_em_suppression}(c). }
	\label{fig:MD17002_fvar_emiss_suppression}
\end{figure}

In Fig.~\ref{fig:MD17_old} we show the spectral fit of MD18a, now with the reflection split into its \textit{soft} and \textit{hard} components. This fit is distinguished from model \textit{\textbf{2C}} in that systematic errors of 0.5\% on the model are used instead of 1\% on the data in the PCA bandpass. This yields a poor fit quality, ($\chi^2_{\nu} = 254.6/91$), however it is clear that even though this difference in systematic errors is fairly minor, the obtained fit is drastically different. We also show the best energy-dependent timing fit for this spectrum in Fig.~\ref{fig:MD17002} with turbulence, emissivity and suppression profiles in Fig.~\ref{fig:MD17002_fvar_emiss_suppression}.

\section{A Developed Physical Picture for the Observed Time Lags}
\label{PhysicalLags}

\begin{figure}
	\includegraphics[width=\columnwidth]{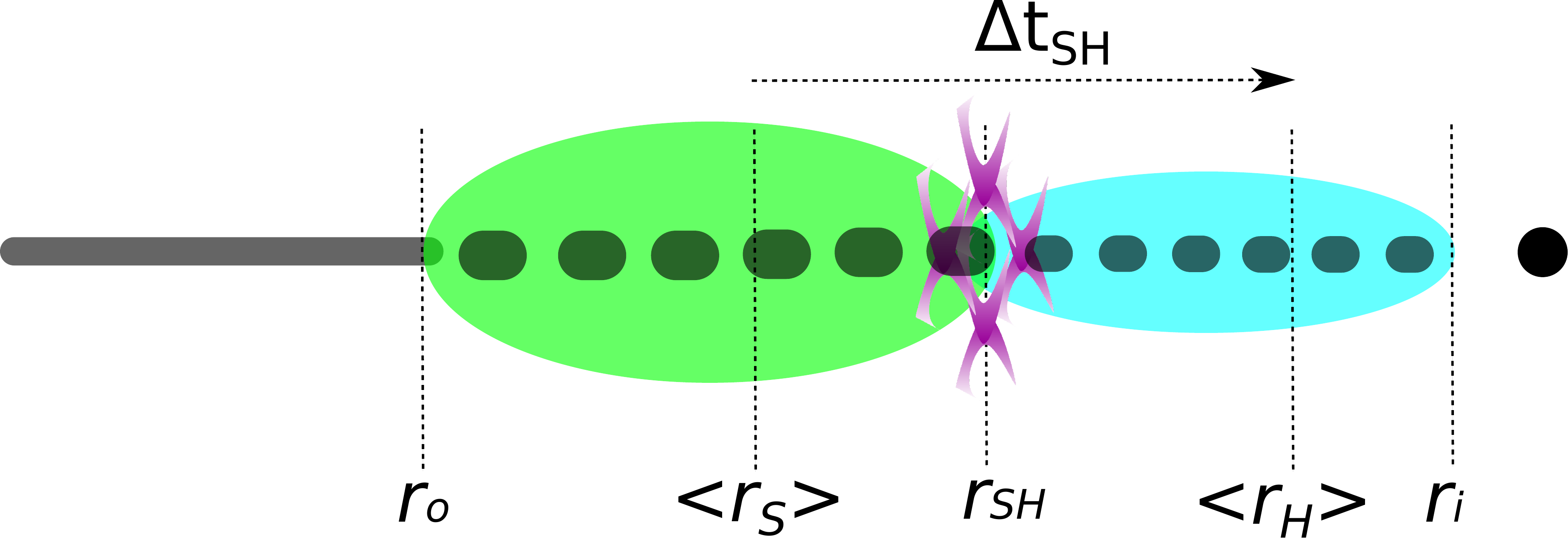}%
	\caption{Simplified picture of the lag origin in the two Compton component model. Colours as in Fig~\ref{fig:TwoGeometry}.}
	\label{fig:TwoLag}
\end{figure}
In MD18a, we demonstrated how the maximal time lag - that for the lowest frequencies generated at $r_o$ - can be understood in terms of fluctuations propagating from the characteristic \textit{soft} to the characteristic \textit{hard} position, $<r_S>$ and $<r_H>$. These are simply the emissivity weighted average positions within each region,
\vspace*{-2pt}
\begin{equation}
\label{eq:raves}
\left<r_S\right>=\frac{\int_{r_{SH}}^{r_o}r^2 \epsilon(r)dr}{\int_{r_{SH}}^{r_o}r \epsilon(r)dr},\,\,\,\ \left<r_H\right>=\frac{\int_{r_i}^{r_{SH}}r^2 \epsilon(r)dr}{\int_{r_i}^{r_{SH}}r \epsilon(r)dr}.
\end{equation}
The maximum raw lag is then the propagation time between these radii
\begin{equation}
\label{eq:lagmax}
\begin{aligned}
\Delta t_{SH} &= \int_{\left<r_H\right>}^{\left<r_S\right>} \frac{dr}{rf_{visc}(r)} \\
&= \frac{2\pi R_g}{Bc}\left[\frac{\left<r_S\right>^{m+3/2}}{m+3/2} - \frac{ \left<r_H\right>^{m+3/2}}{m+3/2} + \frac{\left<r_S\right>^m}{m} -\frac{\left<r_H\right>^m}{m}\right].
\end{aligned}
\end{equation}
This raw lag is illustrated schematically in Fig.~\ref{fig:TwoLag}. The lag we measure is then diluted due to the overlap of the \textit{soft} spectral component into the High band and the \textit{hard} component into the Low band (\citealt{U14}), yielding
\begin{multline}
\label{eq:tau_diluted}
tan[2\pi f_o \tau_{dil}]= \\
\frac{\text{sin}(2\pi f_o \Delta t_{SH})(F_{\textit{soft}}^L F_{\textit{hard}}^H-F_{\textit{soft}}^H F_{\textit{hard}}^L)}{F_{\textit{soft}}^L F_{\textit{soft}}^H+F_{\textit{hard}}^L F_{\textit{hard}}^H + \text{cos}(2\pi f_o \Delta t_{SH})[F_{\textit{soft}}^L F_{\textit{hard}}^H+F_{\textit{soft}}^H F_{\textit{hard}}^L]},
\end{multline}
where $F_{i}^{j}$ is the total flux due to spectral component $i$ in band $j$. For sufficiently `peaked' emissivity profiles, (i.e. the emission profile is approximately delta functions at $<r_S>$ and $<r_H>$ and negligible elsewhere), $f_o$ can be replaced with $f$, generalising equation~(\ref{eq:tau_diluted}) to all frequencies. Introducing the \textit{soft}-\textit{hard} damping mechanism, frequencies $f<f_{visc}(r_{SH})$ are also suppressed by a factor $D_{SH}$ in the \textit{hard} region relative to the \textit{soft}, and so we arrive at a modified form for the final measured lag, 
\begin{multline}
\label{eq:tau_physical}
tan[2\pi f \tau_{fin}(f)]= \\
\frac{\text{sin}(2\pi f \Delta t_{SH})(F_{\textit{soft}}^L F_{\textit{hard}}^H-F_{\textit{soft}}^H F_{\textit{hard}}^L)}{D_{SH}F_{\textit{soft}}^L F_{\textit{soft}}^H+F_{\textit{hard}}^L F_{\textit{hard}}^H + \text{cos}(2\pi f \Delta t_{SH})[F_{\textit{soft}}^L F_{\textit{hard}}^H+F_{\textit{soft}}^H F_{\textit{hard}}^L]}.
\end{multline}

\begin{figure}
	\includegraphics[width=\columnwidth]{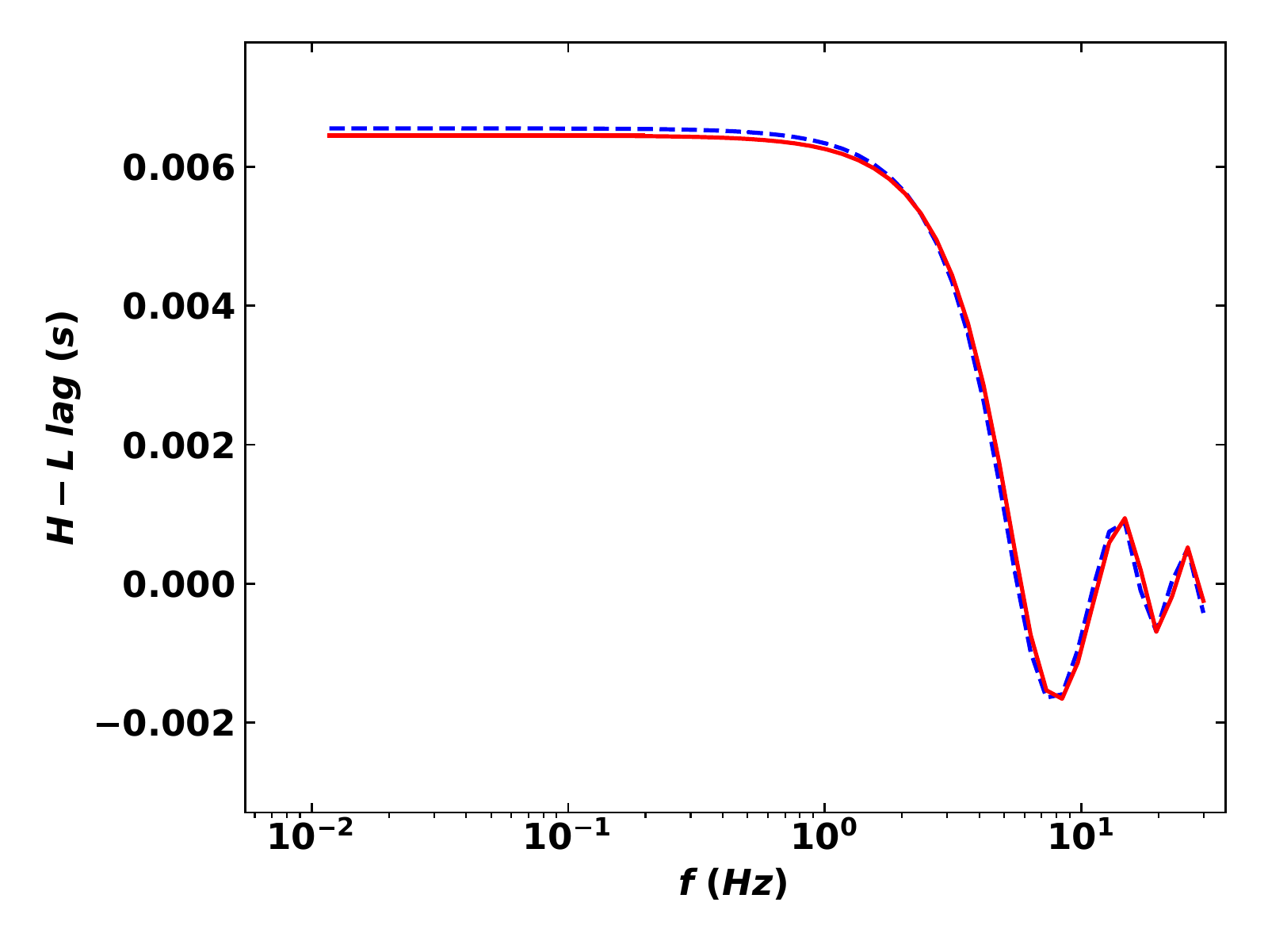}
	\caption{Comparison of lags predicted by the analytic model adapted from IvdK13 (equation~\ref{eq:tau_analytic}, blue dashed line) and the physically-motivated lag model of equation~(\ref{eq:tau_physical}) (red solid line) for simple delta function emissivities at arbitrary $<r_S>$ and $<r_H>$. Here we use an arbitrarily selected $D_{SH}=3$, viscosity parameters of $B=0.03$ and $m=0.5$ and spectral contributions to the Low and High bands from Fig.~\ref{fig:MD17_old}.}
	\label{fig:SimpleHLLag}
\end{figure}

Equation~(\ref{eq:tau_physical}) holds well at low frequencies in all cases, with only minor discrepancies introduced by the width of the Fourier-space generator Lorentzians away from $<r_S>$ and $<r_H>$. For peaked emissivities, the lags at other frequencies are also consistent with equation~(\ref{eq:tau_physical}). In Fig.~~\ref{fig:SimpleHLLag} we demonstrate this by comparing the lags of equation~(\ref{eq:tau_physical}) with the rigorous prediction of equation~(\ref{eq:tau_analytic}) for delta function emissivities at arbitrary $<r_S>$ and $<r_H>$. We note that the models agree remarkably well, showing that lag terms of higher order than equation~(\ref{eq:tau_physical}) can be ignored in the case of this very simple emissivity.

However for broader emission profiles, the lag can significantly diverge from the equation~(\ref{eq:tau_analytic}) prediction, since this breaks the key assumption in equation~(\ref{eq:lagmax}) of the emission being dominated by $<r_S>$ and $<r_H>$. Ultimately the continuum of emission leads to a greater contribution from the lags associated with other annuli and cross-term interference, and equation~(\ref{eq:tau_physical}) breaks down. The broader the emission profile, the more this physically simplified result diverges from the complex analytic prediction of equations~(\ref{eq:Complex_Cross_Spectrum})-(\ref{eq:tau_analytic}).

%%%%%%%%%%%%%%%%%%%%%%%%%%%%%%%%%%%%%%%%%%%%%%%%%%
\section{Uncertainties on Fit Parameters}
\label{Appendix_Uncertainties}
\begin{figure}
	\includegraphics[width=\columnwidth]{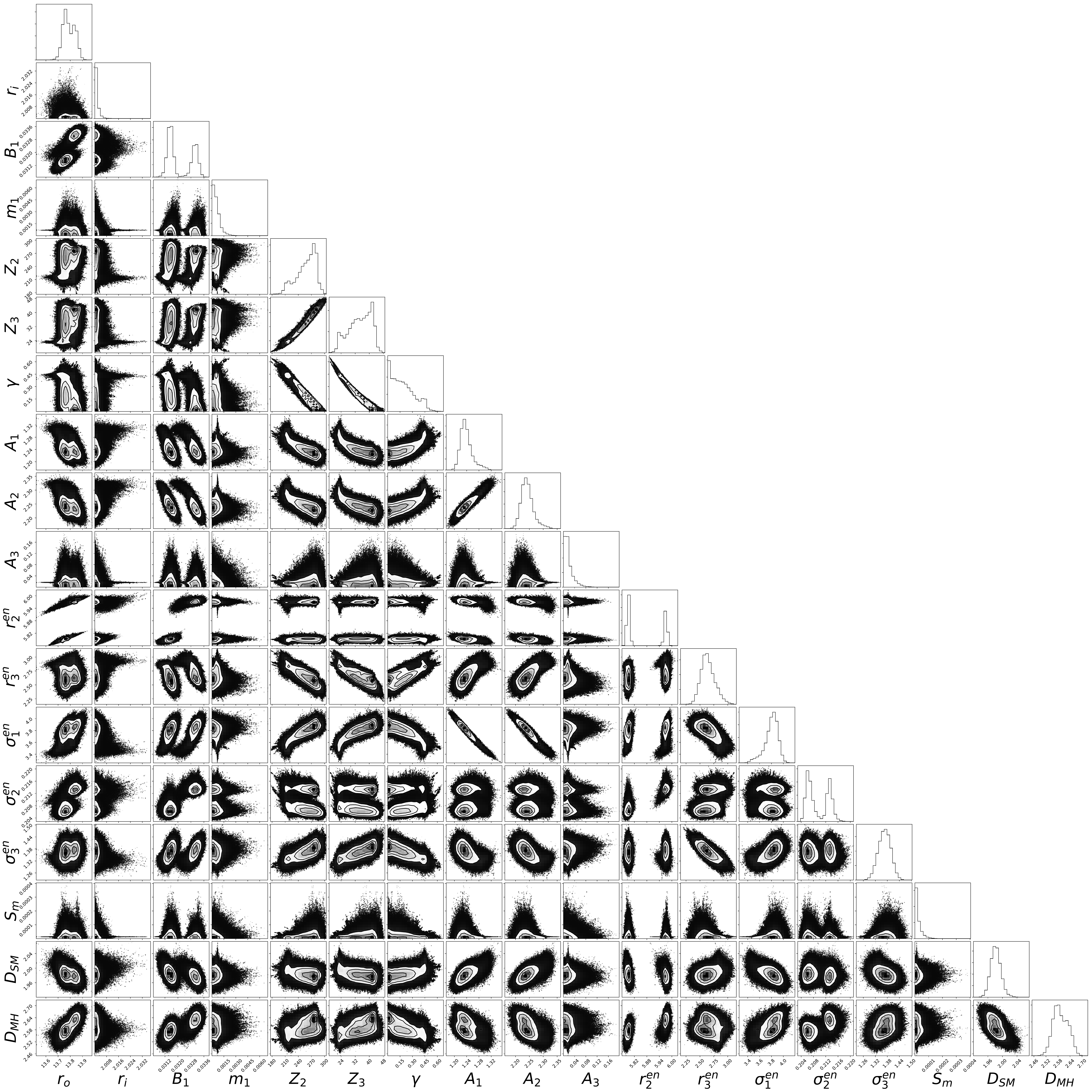}
	\caption{Posterior probability distributions found from the MCMC fit of Fig.~\ref{fig:J2003}, shown as an example of the parameter space degeneracies. Interior contour plots show the optimal regions of parameter space for all pairs of parameters. Panels at the top of each column show the 1-dimensional posterior distributions of each parameter, integrated over all other parameters. Note that this plot is best viewed digitally to avoid pixelization of the contours.}
	\label{fig:CornerPlot}
\end{figure}

Due to its sheer size, the parameter space formed by the final model in this work suffers from some inherent degeneracy. Several of the posterior distributions obtained for these parameters from the MCMC chains were therefore non-gaussian. Partly for this reason we do not quote parameters with errors, but instead show an example corner plot of the posterior distributions for a spectral-timing fit in Fig.~\ref{fig:CornerPlot}. This figure shows the posterior distributions for all pairs of parameters for the fit of Fig.~\ref{fig:J2003}. A good example of the `banana' structure characteristic of degeneracy can be see in the $\gamma$-$Z^\epsilon_2$ or $\gamma$-$Z^\epsilon_3$ distributions in Fig.~\ref{fig:CornerPlot}. This is clearly because the parameters $\gamma$, $Z_\epsilon^1$ and $Z^\epsilon_2$ could be dependently varied to form similar $\epsilon(r)$ profiles. Also, bimodality is exhibited by several  parameters (e.g. $B_1$, $r_2^{en}$), and parameter cross-sections (e.g. $\sigma_{2}^{en}$-$\sigma_{3}^{en}$), making error estimations for these parameters meaningless. By contrast, the damping parameters of $D_{SM}$ and $D_{MH}$ are quite independent of other parameters and could be well determined for a given spectral prior variant. However, even when parameters are well constrained and non-degenerate in the timing fits, these parameters are highly sensitive to the spectral fit used as a prior, and so may vary dramatically for alternative spectral fits.

\onecolumn

\bsp
\label{lastpage}

\begin{thebibliography}{99}
	
\bibitem[\protect\citeauthoryear{Ar\'{e}valo \& Uttley}{2006}]{AU06}
Ar\'{e}valo P., Uttley P., 2006, MNRAS, 367, 801

\bibitem[\protect\citeauthoryear{Arnaud, Borkowski \& Harrington}{1996}]{ABH96}
Arnaud K., Borkowski K.J., Harrington J.P., 1996, ApJ, 462, L75

\bibitem[\protect\citeauthoryear{Axelsson et al.}{2005}]{A05}
Axelsson M., Borgonovo L., Larsson S., 2005, A\&A, 438, 999

\bibitem[\protect\citeauthoryear{Axelsson et al.}{2008}]{A08}
Axelsson M., Hjalmarsdotter L., Borgonovo L., Larsson S., 2008, A\&A, 490, 253

\bibitem[\protect\citeauthoryear{Axelsson \& Done}{2018}]{AD18}
Axelsson M., Done C., 2018, MNRAS, 480 (1), 751

\bibitem[\protect\citeauthoryear{Balbus \& Hawley}{1998}]{BH98}
Balbus S.A., Hawley J.F., 1998, RvMP, 70, 1

\bibitem[\protect\citeauthoryear{Basak \& Zdziarski}{2016}]{BZ16}
Basak R., Zdziarski A.A., 2016, MNRAS, 458, 2199

\bibitem[\protect\citeauthoryear{Basak et al.}{2017}]{B17}
Basak R., Zdziarski A.A., Parker M., Islam N., 2017, MNRAS, 472, 4220

\bibitem[\protect\citeauthoryear{Belloni et al.}{2002}]{B02}
Belloni T., Psaltis D., van der Klis M., 2002, ApJ, 572, 392

\bibitem[\protect\citeauthoryear{Beloborodov}{1999}]{B99}
Beloborodov A.M., 1999, ApJ, 510, L123

\bibitem[\protect\citeauthoryear{Blaes}{2013}]{B13}
Blaes O., 2013, Space Sci. Rev., 103, 21

\bibitem[\protect\citeauthoryear{Churazov, Gilfanov \& Revnivtsev}{2001}]{CGR01}
Churazov E., Gilfanov M., Revnivtsev M., 2001, MNRAS, 321, 759

\bibitem[\protect\citeauthoryear{Di Salvo et al.}{2001}]{DS01}
Di Salvo T., Done C., $\dot{Z}$ycki P.T., Burderi L., Robba N.R., 2001, ApJ, 547, 1024

\bibitem[\protect\citeauthoryear{De Marco et al.}{2015}]{DM15}
De Marco B., Ponti G., Mu\~{n}oz-Darias T., Nandra K., 2015, ApJ, 814, 50

\bibitem[\protect\citeauthoryear{De Marco \& Ponti}{2016}]{DMP16}
De Marco B., Ponti G., 2016, ApJ, 826, 70

\bibitem[\protect\citeauthoryear{De Marco et al.}{2017}]{DM17}
De Marco B., Ponti G., Petrucci P.O. et al., 2017, MNRAS, 471, 1475

\bibitem[\protect\citeauthoryear{Done, Gierli\'{n}ski \& Kubota}{2007}]{DGK07}
Done C., Gierli\'{n}ski M., Kubota A., 2007, A\&ARv, 15, 1 (DGK07)

\bibitem[\protect\citeauthoryear{Esin, McClintock \& Narayan}{1997}]{EMN97}
Esin A.A., McClintock J.E., Narayan R., 1997, ApJ, 489 (2), 865

%\bibitem[\protect\citeauthoryear{Fabian et al.}{2014}]{F14}
%Fabian A.C., Parker M.L., Wilkins D.R.,  et al., 2014, MNRAS, 439, 2307

\bibitem[\protect\citeauthoryear{Foreman-Mackey et al.}{2013}]{DFM13}
Foreman-Mackey D., Hogg D.W., Lang D., Goodman J., 2013, Publ. Astron. Soc. Pac, 125, 306

\bibitem[\protect\citeauthoryear{Fragile et al.}{2007}]{F07}
Fragile P.C., Blaes O.M., Anninos P., Salmonson J.D., 2009, ApJ, 668, 417

\bibitem[\protect\citeauthoryear{Gardner \& Done}{2014}]{GD14}
Gardner E., Done C., 2014, MNRAS, 442, 2456

\bibitem[\protect\citeauthoryear{Generozov et al.}{2014}]{GBFH14}
Generozov A., Blaes O., Fragile P.C., Henisey K.B., 2014, ApJ, 780, 81

\bibitem[\protect\citeauthoryear{Gierli{\'n}ski et al.}{1997}]{G97}
Gierli{\'n}ski M., Zdziarski A.A., Done C. et al., 1997, MNRAS, 288, 958

\bibitem[\protect\citeauthoryear{Gierli{\'n}ski, Done \& Page}{2009}]{GDP09}
Gierli{\'n}ski M., Done C., Page K., MNRAS, 392, 1106

%\bibitem[\protect\citeauthoryear{Gilfanov, Churazov \& Revnivtsev}{2000}]{GCR00}
%Gilfanov M., Churazov E., Revnivtsev M., 2000, MNRAS, 316, 923

\bibitem[\protect\citeauthoryear{Grinberg et al.}{2014}]{G14}
Grinberg V., Pottshmidt K., B{\"o}ck M. et al., 2014, A\&A, 565, A1

%\bibitem[\protect\citeauthoryear{Heil, Vaughan \& Uttley}{2012}]{HVU12}
%Heil L.M., Vaughan S. \& Uttley P., 2012, MNRAS, 422, 3620

\bibitem[\protect\citeauthoryear{Haardt \& Maraschi}{1993}]{HM93}
Haardt F., Maraschi L., 1993, ApJ, 413, 507

\bibitem[\protect\citeauthoryear{Henisey, Blaes \& Fragile}{2012}]{HBF12}
Henisey K.B., Blaes O.M., Fragile P.C., 2012, ApJ, 761, 18

\bibitem[\protect\citeauthoryear{Hogg \& Reynolds}{2017}]{HR17}
Hogg J.D., Reynolds C.S., 2017, ApJ, 834, 80

\bibitem[\protect\citeauthoryear{Ibragimov et al.}{2005}]{I05}
Ibragimov A., Poutanen J., Gilfanov M., Zdziarski A.A. \& Shrader C.R., 2005, MNRAS, 362, 1435

\bibitem[\protect\citeauthoryear{Ingram et al.}{2009}]{IDF09}
Ingram A., Done C., Fragile P.C., 2009, MNRAS, 397 (1), L101

\bibitem[\protect\citeauthoryear{Ingram \& Done}{2011}]{ID11}
Ingram A., Done C., 2011, MNRAS, 415 (3), 2323 (ID11)

\bibitem[\protect\citeauthoryear{Ingram \& Done}{2012a}]{ID12a}
Ingram A., Done C., 2012, MNRAS, 419, 2369

\bibitem[\protect\citeauthoryear{Ingram \& Done}{2012b}]{ID12b}
Ingram A., Done C., 2012, MNRAS, 427, 934

\bibitem[\protect\citeauthoryear{Ingram \& van der Klis}{2013}]{IvdK13}
Ingram A., van der Klis M., 2013, MNRAS, 434, 1476 (IvdK13)

\bibitem[\protect\citeauthoryear{Ingram et al.}{2016}]{I16}
Ingram A., van der Klis M., Middleton M. et al., 2016, MNRAS, 461 (2), 1967

\bibitem[\protect\citeauthoryear{Kawano et al.}{2017}]{K17}
Kawano T., Done C., Yamada S., Takahashi H., Axelsson M., Fukuzawa Y., 2017, PASJ, 69(2), 36

\bibitem[\protect\citeauthoryear{Klein-Wolt \& van der Klis}{2008}]{KWvdK08}
Klein-Wolt M., van der Klis M., 2008, ApJ, 675, 1407

\bibitem[\protect\citeauthoryear{Kolehmainen, Done \& Diaz Trigo}{2014}]{KDD14}
Kolehmainen M., Done C., Diaz Trigo M., 2014, MNRAS, 437, 613

\bibitem[\protect\citeauthoryear{Kotov, Churazov \& Gilfanov}{2001}]{KCG01}
Kotov O., Churazov E., Gilfanov M., 2001, MNRAS, 327, 799

%\bibitem[\protect\citeauthoryear{Liska et al.}{2017}]{L17}
%Liska M., Hesp C., Tchekhovskoy A., Ingram A., van der Klis M., Markoff S., 2017, MNRAS, 474, L81

\bibitem[\protect\citeauthoryear{Lubow, Ogilvie \& Pringle}{2002}]{LOP02}
Lubow S.H., Ogilvie G.I., Pringle J.E., 2002, MNRAS, 337, 706

\bibitem[\protect\citeauthoryear{Lyubarskii}{1997}]{L97}
Lyubarskii Y.E., 1997, MNRAS, 292, 679

\bibitem[\protect\citeauthoryear{Mahmoud \& Done}{2018}]{MD18a}
Mahmoud R.D., Done C., 2018, MNRAS, 473, 2084 (MD18a)

\bibitem[\protect\citeauthoryear{Makishima et al.}{2008}]{M08}
Makishima K., Takahashi H., Yamada S. et al., 2008, PASJ, 60, 585

\bibitem[\protect\citeauthoryear{Mastroserio, Ingram \& van der Klis}{2018}]{MIvdK18}
Mastroserio G., Ingram A., van der Klis M., 2018, MNRAS, 475, 4027 

%\bibitem[\protect\citeauthoryear{Markoff, Nowak \& Wilms}{2005}]{MNW05}
%Markoff S., Nowak M.A. \& Wilms J., 2005, ApJ, 635, 1203

\bibitem[\protect\citeauthoryear{Misra et al.}{2017}]{M17}
Misra R., Yadav J.S., Chauhan J.V. et al., 2017, ApJ, 835 (2), 195

\bibitem[\protect\citeauthoryear{Miyamoto \& Kitamoto}{1989}]{MK89}
Miyamoto A., Kitamoto S., 1989, Nature, 342, 773

%\bibitem[\protect\citeauthoryear{Mu\~{n}oz-Darias, Motta \& Belloni}{2011}]{MMB11}
%Mu\~{n}oz-Darias T., Motta S. \& Belloni T.M., 2011, MNRAS, 410, 679 

\bibitem[\protect\citeauthoryear{Mushtukov, Ingram \& van der Klis}{2017}]{MIvdK17}
Mushtukov A.A., Ingram A., van der Klis M., 2017, MNRAS, 474, 2259

%\bibitem[\protect\citeauthoryear{Narayan, Kato \& Honma}{1997}]{NKH97}
%Narayan R., Kato S. \& Honma F., 1997, ApJ, 476, 49

\bibitem[\protect\citeauthoryear{Narayan \& Yi}{1995}]{NY95}
Narayan R., Yi I., 1995, ApJ, 452, 710

\bibitem[\protect\citeauthoryear{Noble \& Krolik}{2009}]{NK09}
Noble S.C., Krolik J.H., 2009, ApJ, 703, 964

\bibitem[\protect\citeauthoryear{Nowak \text{et al.}}{1999}]{N99}
Nowak M.A., Vaughan B.A., Wilms J., Dove J.B., Begelman M.C., 1999, ApJ, 510, 874

\bibitem[\protect\citeauthoryear{Nowak}{2000}]{N00}
Nowak M.A., 2000, MNRAS, 318, 361

\bibitem[\protect\citeauthoryear{Papadakis \& Lawrence}{1993}]{PL93}
Papadakis I.E., Lawrence A., 1993, MNRAS, 261, 612

%\bibitem[\protect\citeauthoryear{Nowak et al.}{2011}]{N11}
%Nowak M.A., Hanke M., Trowbridge S.N.,  et al., 2011, ApJ, 728, 13

\bibitem[\protect\citeauthoryear{Pottschmidt et al.}{2003}]{P03}
Pottschmidt K., Wilms J., Nowak M.A. et al., 2003, A\&A, 407, 1039

%\bibitem[\protect\citeauthoryear{Poutanen \& Coppi}{1998}]{PC98}
%Poutanen J. \& Coppi P.S., 1998, Phys. Scripta, T77, 57

%\bibitem[\protect\citeauthoryear{Poutanen \& Vurm}{2009}]{PV09}
%Poutanen J. \& Vurm I., 2009, ApJ, 690, L97

%\bibitem[\protect\citeauthoryear{Poutanen \& Veledina}{2014}]{PV14}
%Poutanen J. \& Veledina A., 2014, Space Sci. Rev., 183, 61

\bibitem[\protect\citeauthoryear{Poutanen, Veledina \& Zdziarski}{2018}]{PVZ18}
Poutanen J., Veledina A., Zdziarski A.A., 2018, A\&A, 614, A79

%\bibitem[\protect\citeauthoryear{Rapisarda, Ingram \& van der Klis}{2014}]{RIvdK14}
%Rapisarda S., Ingram A. \& van der Klis M., 2014, MNRAS, 440, 2882

\bibitem[\protect\citeauthoryear{Rapisarda et al.}{2016}]{R16}
Rapisarda S., Ingram A., Kalamkar M., van der Klis M., 2016, MNRAS, 462, 4078

\bibitem[\protect\citeauthoryear{Rapisarda, Ingram \& van der Klis}{2017a}]{R17a}
Rapisarda S., Ingram A., van der Klis M., 2017, MNRAS, 469 (2), 2017 (R17a)

\bibitem[\protect\citeauthoryear{Rapisarda, Ingram \& van der Klis}{2017b}]{R17b}
Rapisarda S., Ingram A., van der Klis M., 2017, 472, 3821 (R17b)

%\bibitem[\protect\citeauthoryear{Reis, Fabian \& Miller}{2010}]{RFM10}
%Reis R.C, Fabian A.C \& Miller J.M., 2010, MNRAS, 402, 836

%\bibitem[\protect\citeauthoryear{Remillard \& McClintock}{2006}]{RM06}
%Remillard \& McClintock, 2006, in \textit{Compact Stellar X-Ray Sources}, Ch. 4, Cambridge University Press, ed. Lewin, W.H.G. \& van der Klis, M.

%\bibitem[\protect\citeauthoryear{Revnivtsev, Gilfanov \& Churazov}{1999}]{RGC99}
%Revnivtsev M., Gilfanov M. \& Churazov E., 1999, A\&A, 347, L23

%\bibitem[\protect\citeauthoryear{Revnivtsev et al.}{2011}]{R11}
%Revnivtsev M., Potter S., Kniazev A., Burenin R., Buckley D.A.H. \& Churazov E., 2011, MNRAS, 411, 1317

%\bibitem[\protect\citeauthoryear{Rykoff et al.}{2007}]{Ry07}
%Rykoff, E.S., Miller J.M, Steeghs D., Torres M.A.P., 2007, ApJ, 666, 1129

\bibitem[\protect\citeauthoryear{Shakura \& Sunyaev}{1973}]{SS73}
Shakura N.I., Sunyaev R.A., 1973, A\&A, 24, 337

\bibitem[\protect\citeauthoryear{Tomsick et al.}{2014}]{T14}
Tomsick J.A. et al., 2014, ApJ, 780, 78

\bibitem[\protect\citeauthoryear{Timmer \& K{\"o}nig}{1995}]{TK95}
Timmer J., K{\"o}nig M., 1995, A\&A, 300, 707

\bibitem[\protect\citeauthoryear{Torii et al.}{2011}]{T11}
Torii S., Yamada S., Makishima K, et al., 2011, PASJ, 63, S771

\bibitem[\protect\citeauthoryear{Uttley et al.}{2011}]{U11}
Uttley P., Wilkinson T., Cassatella P., Wilms E., Pottschmidt K., Hanke M., B{\"o}ck M., 2011, MNRAS, 414, L60

\bibitem[\protect\citeauthoryear{Uttley et al.}{2014}]{U14}
Uttley P., Cackett E.M., Fabian A.C., Kara E., Wilkins D.R., 2014, Astron. Astrophys. Rev., 22, 72

\bibitem[\protect\citeauthoryear{van der Klis}{1989}]{vdK89}
van der Klis M., 1989, in \textit{Timing Neutron Stars: proceedings of the NATO Advanced Study Institute on Timing Neutron Stars}, p.27, Kluwer Academic / Plenum Publishers, ed. {\"O}gelman H. \& van den Heuvel E.P.J.

\bibitem[\protect\citeauthoryear{Veledina}{2016}]{V16}
Veledina A., 2016, ApJ, 832, 181
                           
\bibitem[\protect\citeauthoryear{Wijnands \& van der Klis}{1999}]{WvdK99}
Wijnands R., van der Klis M., 1999, ApJ, 522 (2), 965

\bibitem[\protect\citeauthoryear{Wilkinson \& Uttley}{2009}]{WU09}
Wilkinson T., Uttley P., 2009, MNRAS, 397, 666

\bibitem[\protect\citeauthoryear{Wilms, Allen \& McCray}{2000}]{WAM00}
Wilms J., Allen A., McCray R. 2000, ApJ, 542, 914

\bibitem[\protect\citeauthoryear{Yamada et al.}{2013}]{Y13}
Yamada S., Makishima K., Done C., Torii S., Noda H., Sakurai S., 2013, PASJ, 65, 80

\bibitem[\protect\citeauthoryear{Zdziarski, Johnson \& Magdziarz}{1996}]{ZJM96}
Zdziarski A.A., Johnson W.N., Magdziarz P., 1996, MNRAS, 283, 193

\end{thebibliography}
\end{document}